\documentclass{aastex6}

\newcommand{\kzz}{K$_{zz}$ }

\begin{document}


\title{Microphysics of KCl and ZnS Clouds on GJ 1214 b}


\author{Peter Gao\altaffilmark{1,2}}
\affil{University of California, Berkeley \\
Berkeley, CA 94720, USA}

\and

\author{Bj\"orn Benneke}
\affil{Universit\'e de Montr\'eal \\
2900 Edouard Montpetit Blvd. \\
Montreal, QC H3T 1J4, Canada}




\altaffiltext{1}{51 Pegasi b Fellow}
\altaffiltext{2}{gaopeter@berkeley.edu}

\begin{abstract}

Clouds in the atmospheres of exoplanets confound characterization efforts by reducing, eliminating, and distorting spectral signatures of molecular abundances. As such, interpretations of exoplanet spectra strongly depend on the choice of cloud model, many of which are highly simplified and lack predictive power. In this work, we use a cloud model that treat microphysical processes to simulate potassium chloride (KCl) and zinc sulfide (ZnS) clouds in the atmosphere of the super Earth GJ 1214 b and how they vary as a function of the strength of vertical mixing and the atmospheric metallicity. Microphysical processes control the size and spatial distribution of cloud particles, allowing for the computation of more physical cloud distributions than simpler models. We find that the mass and opacity of KCl clouds increase with mixing strength and metallicity, with the particle size and spatial distribution defined by nucleation, condensation, evaporation, and transport timescales. ZnS clouds cannot form without the presence of condensation nuclei, while heterogeneous nucleation of ZnS on KCl reduces particle sizes compared to pure KCl cases. In order to explain the flat transmission spectrum of GJ 1214b with homogeneously nucleated KCl clouds, the atmospheric metallicity must be at least 1000 $\times$ solar, and the eddy diffusivity must be at least 10$^{10}$ cm$^2$ s$^{-1}$. We predict that JWST observations of GJ 1214 b may reveal the presence of methane, carbon monoxide, and water, allowing for constraints to be placed on atmospheric metallicity and C/O ratio.

\end{abstract}




\section{Introduction} \label{sec:intro}


Clouds and hazes are prevalent in exoplanet atmospheres, where they often impede probing of atmospheric composition \citep[e.g.][]{gibson2012,gibson2013,deming2013,jordan2013,mandell2013,sing2013,chen2014,schlawin2014,wilkins2014,fukui2014,mallonn2016,damiano2017,kreidberg2018}. Their presence is typically revealed through diminished amplitudes of spectral features in the transmission spectrum, as optically thick cloud decks block stellar photons from reaching to depths below. This phenomenon is observed across many exoplanets of various sizes, effective temperatures, and stellar irradiation levels, with large variations in ``cloudiness'' between otherwise similar exoplanets \citep[e.g.][]{crossfield2013,kreidberg2014,knutson2014a,knutson2014b,fraine2014,sing2016,iyer2016,bruno2018}.

There are now some hints of correlations between cloudiness and other planetary properties, such as equilibrium temperature and irradiation \citep{stevenson2016,heng2016,fu2017}, but the processes that control the formation and distribution of exoplanet clouds are still mostly unknown. This is a profound problem, as clouds are strongly coupled to the atmospheric radiation environment, chemical composition, and dynamics. For example, the presence of a cloud formed from condensation gives clues to the temperature structure and metallicity of an atmosphere, as cloud formation requires a supersaturation of the condensing species. In addition, clouds obfuscate investigations of exoplanet atmospheres by hiding molecular spectral features, and thus future exoplanet observing missions and programs would greatly benefit from a theoretical framework that can determine \textit{a priori} whether an exoplanet is cloudy or clear. 

Previous efforts in detailed modeling of exoplanet clouds can be split into three categories: Equilibrium condensation, grain chemistry, and more recently, collisional growth. Equilibrium condensation presumes an atmosphere with a composition determined by thermochemical equilibrium, and that certain species can be in their condensed phase if it is energetically favorable. The resulting condensed material are distributed vertically such that the sedimentation of cloud particles is balanced by vertical lofting due to eddy mixing \citep{ackerman2001}. Using this framework, \citet{morley2012} showed that the formation of sulfide clouds (e.g. Na$_{2}$S) could explain the infrared colors of cooler brown dwarfs, while \citet{morley2013,morley2015} was able to explain the flat transmission spectrum of the super Earth GJ 1214 b by appealing to the formation of KCl and ZnS clouds in a high (100--1000 $\times$ solar) metallicity atmosphere. \citet{charnay2015a,charnay2015b} extended this scheme to 3D and found that submicron KCl and ZnS particles can be lofted to high enough altitudes such that the resulting transmission spectrum is flat within the uncertainties of the data. Similarly, \citet{parmentier2016} applied equilibrium condensation in 3D to hot Jupiters and showed how transitions in cloud composition with effective temperature could explain the variations in exoplanet light curves observed by Kepler. However, while these models have been successful in explaining the available data, they do not take into account the physical processes that govern particle sizes and number densities, namely microphysical processes. This deficiency dampens their predictive powers.    

Grain chemistry models assume that cloud formation is a kinetic process, characterized by the growth and evaporation of cloud particles with mixed compositions via heterogeneous chemical reactions on their surfaces. This framework has been developed in great detail for brown dwarf and hot Jupiter atmospheres, where the cloud formation processes is assumed to begin with the nucleation of TiO$_{2}$ clusters in the upper atmosphere that are then transported downwards and act as nucleation sites for an array of condensing species, including MgSiO$_{3}$, Mg$_{2}$SiO$_{4}$, SiO$_{2}$, Al$_{2}$O$_{3}$, and Fe \citep{helling2001,woitke2003,woitke2004,helling2004,helling2006,helling2008a,helling2008b,witte2009,witte2011,lee2015,lee2015b,helling2016}. These models have been compared to brown dwarf emission spectra \citep{witte2011}, and have shown that a vertical gradient in cloud composition is likely in brown dwarf and hot exoplanet atmospheres. These models have also been extended to 3D \citep{lee2015,helling2016,lines2018}, and showed that HD 189733b and HD 209458b are enveloped by mineral clouds with vertical and latitudinal variations in composition due to dynamics and global temperature differences, with the former planet exhibiting a lower cloud deck, explaining the more pronounced molecular features in its transmission spectra. Though these models have greatly illuminated the complexities of brown dwarf/exoplanet cloud processes, they are difficult to generalize due to their reliance on the specific TiO$_{2}$ nucleation pathway for cloud formation. As a result, they have not yet been used to investigate cooler, smaller exoplanets. 

Collisional growth models assume that aerosol particles grow mainly by colliding with each other and sticking (coagulation) and are typically applied to nonvolatile aerosols in photochemical hazes. These models have been recently applied to giant exoplanets \citep{lavvas2017} and GJ 1214 b \citep{kawashima2018}, and show that the resulting particle distribution is strongly dependent on the photochemical production rate of haze precursors. These models have yet to be applied to volatile species that are thought to compose exoplanet clouds.

In this work, we explore the formation and evolution of exoplanet clouds from a perspective especially applicable to cooler atmospheres: Cloud microphysics. On Earth, water clouds form through nucleation of preexisting water molecules onto condensation nuclei (heterogeneous nucleation) or, less frequently, by homogeneous nucleation where no seeds are necessary; these cloud particles can then grow by condensation and coagulation, or shrink by evaporation, and are transported through sedimentation, advection, and turbulent mixing. These processes are likely at work across the Solar System, controlling the sulfuric acid clouds of Venus \citep{james1997,gao2014,mcgouldrick2017}, the CO$_{2}$ and water clouds of Mars \citep{michelangeli1993,colaprete1999}, the ammonia clouds of Jupiter and Saturn \citep{rossow1978,carlson1988,ohno2017}, and the hydrocarbon clouds of Titan, Uranus, and Neptune \citep{moses1992,barth2003,barth2004,barth2006,lavvas2011}. Cloud microphysics is a variation on grain chemistry in that it also treats cloud formation as a kinetics process, but it presumes that the condensing material already exists as free floating molecules, whereas in grain chemistry many of the condensed species form on the grains themselves.    

We use the 1-dimensional Community Aerosol and Radiation Model for Atmospheres (CARMA) to simulate KCl and ZnS clouds on GJ 1214 b, a warm ``super Earth'' with a flat transmission spectrum measured with high precision \citep{gillon2014,kreidberg2014}. These observations potentially allow for tight constraints to be placed on cloud and related atmospheric properties, such as atmospheric metallicity and turbulent mixing strength. In addition, we can also investigate the assumptions and results of previous studies that have targeted this world that use simpler cloud models \citep{morley2013,morley2015,charnay2015a,charnay2015b}. Our study complements the recent work by \citet{ohno2018}, who also investigated clouds in the atmosphere of GJ 1214 b using cloud microphysics. Our approaches differ through our choice of cloud formation pathway and treatments of the particle size distribution, and will be compared in ${\S}$\ref{sec:discussion5}.

In ${\S}$\ref{sec:model5}, we briefly describe our cloud microphysics model, including the augmentations needed to simulate exoplanet clouds. We discuss our results in ${\S}$\ref{sec:results5}, where we show how the cloud distribution varies with eddy mixing, metallicity, and formation pathway, and how our results compare to observations. We discuss the implications of our results and make recommendations for future investigations in ${\S}$\ref{sec:discussion5}. We present our conclusions in ${\S}$\ref{sec:conclusions5}. 

\section{Model}\label{sec:model5}

\subsection{Model Description}\label{sec:modeldes}

CARMA is a 1--dimensional Eulerian forward model that solves the discretized continuity equation for aerosol particles subject to vertical transport and production and loss due to particle nucleation (homogeneous and heterogeneous), condensation, evaporation, and coagulation \citep{turco1979,toon1988,jacobson1994,ackerman1995,bardeen2008}. Starting from an initial state, CARMA time steps towards an equilibrium particle distribution by balancing the rates of the aforementioned processes. In addition, CARMA resolves the particle size distribution using mass bins, thus avoiding assumptions of analytical particle size distributions (e.g. lognormal), which may lead to errors when the actual size distribution is irregular. CARMA has recently been applied to hot Jupiters for the first time \citep{powell2018}, and this study continues this effort for cooler exoplanets. We refer the reader to the appendix of \citet{gao2018} for a complete description of the physics contained in CARMA. 

\begin{figure}[hbt!]
\centering
\includegraphics[width=0.6 \textwidth]{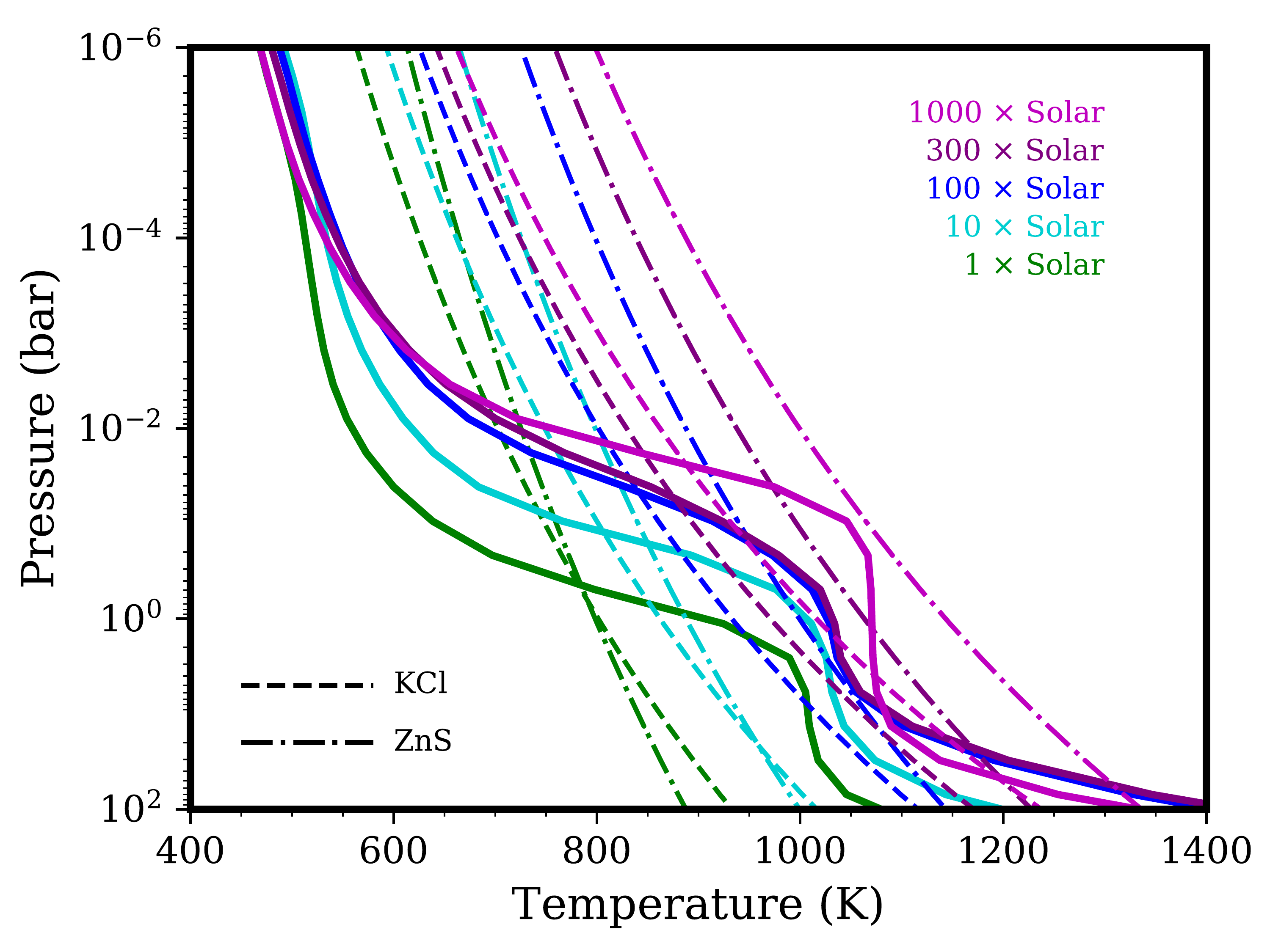}
\caption{Cloudless atmospheric pressure--temperature profiles of GJ 1214 b with 1 $\times$ (green), 10 $\times$ (cyan), 100 $\times$ (blue), 300 $\times$ (purple), and 1000 $\times$ (magenta) solar metallicity. The condensation curves of KCl (dashed) and ZnS (dash-dot) at the same metallicities are included for comparison \citep{morley2012}.}
\label{fig:gj1214btp}
\end{figure}

To investigate the possible cloud distributions in the atmosphere of GJ 1214b using CARMA, we conduct a parameter space exploration by varying the atmospheric metallicity and eddy diffusion coefficient \kzz. The atmospheric metallicity affects cloud distributions by determining the abundance of condensate vapor available to form clouds. GJ 1214b's atmospheric metallicity is currently unknown due to the lack of molecular features in its transmission spectrum \citep{kreidberg2014}, though previous modeling efforts showed that metallicities of 100--1000 $\times$ solar are needed to match said lack of features \citep{morley2015,charnay2015b}. Such high metallicities are consistent with the theoretical mass--metallicity relation for exoplanets \citep{fortney2013}, but this relationship has not been benchmarked in the mass range occupied by GJ 1214b due to lack of observations. In addition, examples exist of exoplanets with metallicities both lower and higher than expected given their masses \citep[e.g.][]{wakeford2017,wakeford2018}. Therefore, we expand our parameter space exploration to include lower metallicities. In all, we explore the cloud distribution at metallicities of 1 $\times$, 10 $\times$, 100 $\times$, 300 $\times$, and 1000 $\times$ solar.

We use the SCARLET atmosphere modeling framework \citep{benneke2013,benneke2015} to generate pressure–temperature (PT) profiles for the atmosphere of GJ 1214b at the given metallicities (Figure \ref{fig:gj1214btp}). These PT profiles are constructed self-consistently assuming radiative–convective–thermochemical equilibrium, line-by-line molecular opacities, and solar C/O ratio. Scenarios with higher metallicity result in higher temperatures for a given pressure due to the increased molecular opacity in the atmosphere. At the temperatures considered and the pressures probed by transmission spectroscopy, KCl and ZnS are the likely condensates, as determined by equilibrium chemistry \citep{lodders1999,morley2013}. Figure \ref{fig:gj1214btp} shows the condensation curves of KCl and ZnS, constructed from their saturation vapor pressures and vapor abundances (see ${\S}$\ref{sec:micro}), crossing the PT profiles of GJ 1214b at several hundred mbar at solar metallicity. At higher metallicities, the condensation curves can cross the PT profile twice, once at depth, and once higher up at pressures lower than 100 mbar. This may result in formation of condensed material at depth in a ``cold trap'' that could reduce the cloud mass at lower pressures \citep[e.g.][]{parmentier2016,powell2018}.

Eddy diffusion is a parameterization of large scale transport and mixing in 1D atmospheric models and affects the cloud distribution by lofting cloud particles to lower atmospheric pressure levels and by upwelling condensate vapor from depth into the cloud formation region. The value of \kzz is uncertain, as its physical basis is not well understood, especially in non-convective regions of the atmosphere. \citet{charnay2015a} computed an effective \kzz for GJ 1214b using 3D GCM simulations that is a power law in pressure, with values that are metallicity dependent, ranging from 7 $\times$ 10$^{6}$ to 3 $\times$ 10$^{7}$ cm$^{2}$ s$^{-1}$ at 1 bar, rising to $\sim$10$^9$ at 1 mbar. \citet{morley2015} used mixing length theory to generate similar \kzz profiles. Studies of other exoplanets' atmospheres have considered \kzz values from 10$^{6}$ to $>$10$^{10}$ cm$^{2}$ s$^{-1}$ \citep[e.g.][]{line2010,kopparapu2012,moses2013,parmentier2013,konopacky2013,miguel2014,hu2014,barman2015,venot2015}. In order to better understand the effect of varying \kzz on cloud distributions, we will assume constant \kzz profiles in our atmospheres, and examine a broad range of values within those that have been used in previous works. Specifically, we will conduct simulations using \kzz values of 10$^7$, 10$^8$, 10$^9$, and 10$^{10}$ cm$^2$ s$^{-1}$. 

\begin{deluxetable}{clcc}
\tablecolumns{4}
\tablecaption{Relevant atmosphere and planetary parameters \label{table:atmparam}}
\tablehead{
\colhead{Symbol} & \colhead{Parameter} & \colhead{Value} & \colhead{References}}
\startdata
\nodata & Gravity\tablenotemark{a} (m s$^{-2}$)				&	8.93	&	\citet{charbonneau2009}	\\
$M_a$ & Atmospheric mole. wt.\tablenotemark{b} (g mol$^{-1}$)		&			&						\\
\nodata &\hspace{0.5cm} Solar metallicity 					&	2.32	&	\nodata				\\
\nodata & \hspace{0.5cm} 10 $\times$ Solar metallicity			&	2.45	&	\nodata				\\
\nodata & \hspace{0.5cm} 100 $\times$ Solar metallicity			&	3.68	&	\nodata				\\
\nodata & \hspace{0.5cm} 300 $\times$ Solar metallicity			&	6.14	&	\nodata				\\
\nodata & \hspace{0.5cm} 1000 $\times$ Solar metallicity			&	12.5	&	\nodata				\\
$\eta$ & Atmospheric viscosity\tablenotemark{c} (10$^{-4}$ Poise)		&	1.85 		&	\citet{white1974}			\\
$\kappa$ & Atmospheric thermal conductivity\tablenotemark{d} (10$^{4}$ ergs s$^{-1}$ m$^{-1}$ K$^{-1}$)	&	4.83		&	\citet{lemmon2016}		\\
$C_p$ & Atmospheric heat capacity (10$^{8}$ ergs g$^{-1}$ K$^{-1}$)	&	1.3  	&	 \citet{kataria2015}		\\
\enddata
\tablenotetext{a}{At 0.1 bars. Varies with the inverse square of increasing distance from the center of the planet.}
\tablenotetext{b}{At 1 bar. Changes with pressure level according to atmospheric composition computed from equilibrium chemistry.}
\tablenotetext{c}{At 1 bar. See Eq. \ref{eq:gj1214bvis} for variations with temperature.}
\tablenotetext{d}{At 1 bar. See Eq. \ref{eq:gj1214bthcond} for variations with temperature.}

\end{deluxetable}

Additional atmospheric and planetary properties that are relevant to cloud formation are given in Table \ref{table:atmparam}. The mean molecular weight of the atmosphere is computed from the atmospheric composition determined by equilibrium chemistry, and varies with atmospheric pressure level by $\sim$1--10\%. The viscosity of the atmosphere is a function of temperature $T$ given by the Sutherland equation

\begin{equation}
\label{eq:gj1214bvis}
\eta(Poise) = 8.76 \times 10^{-5} \left ( \frac{293.85 + 72}{T + 72} \right ) \left ( \frac{T}{293.85} \right )^{1.5}
\end{equation}

\noindent where the numerical constants are those for H$_{2}$ taken from \citet{white1974}. The atmospheric viscosity is needed to calculate the sedimentation velocity of cloud particles. 
	         
Condensational growth in CARMA takes into account latent heating of particles due to the addition of vapor molecules. This heat is removed from the particle by conduction. We set the thermal conductivity $\kappa$ of the atmosphere to be that of pure H$_{2}$, given by 

\begin{equation}
\label{eq:gj1214bthcond}
\kappa(ergs\, s^{-1}\, m^{-1}\, K^{-1}) = 7148.57 + 39.12T + 3.1607 \times 10^{-3} T^{2} 
\end{equation}

\noindent where the numerical coefficients are obtained by fitting to data from \citet{lemmon2016}. For the heat capacity of the atmosphere $C_p$, which is used to correct for $\kappa$ in the kinetic limit, we use a constant value of 1.3 $\times$ 10$^{8}$ ergs g$^{-1}$ K$^{-1}$, appropriate for a H$_{2}$--dominated atmosphere \citep{kataria2015}. 

We neglect variations in atmospheric viscosity, heat capacity, and thermal conductivity due to increasing atmospheric metallicity, as how these values vary is unknown. However, comparing that of H$_2$ to that of higher molecular weight gases, such as steam and N$_2$, shows that viscosity can be higher by about a factor of two, while heat capacity and thermal conductivity can be lower by an order of magnitude for higher metallicity atmospheres \citep{rumble2017}. We tested the effect of changing these quantities by the given factors using CARMA, and found that the resulting cloud distributions did not change significantly.


\subsection{Microphysical Properties of KCl \& ZnS}\label{sec:micro}

Within the ranges in pressures and temperatures we consider for GJ 1214b's atmosphere, and under equilibrium conditions, most atomic K is bound in KCl \citep{lodders1999}. KCl gas undergoes phase transition to condensed KCl if its partial pressure is greater than its saturation vapor pressure, given by \citet{morley2012} as

\begin{equation}
\label{eq:kclsvp}
\log{P_s^{KCl}} = 7.611 - 11382/T
\end{equation}

\noindent where P$^{s}_{KCl}$ is in bars and $T$ is in K. In contrast, molecular ZnS is not thought to exist in the gaseous phase. Instead, atomic Zn gas reacts with hydrogen sulfide, the dominant sulfur compound at the relevant pressure and temperatures, to directly form condensed ZnS and hydrogen gas \citep{visscher2006}. A ``thermochemical saturation vapor pressure'' of Zn over ZnS can be defined such that condensed ZnS will form if the partial pressure of Zn gas exceeds said saturation vapor pressure, given by \citet{morley2012} as

\begin{equation}
\label{eq:znssvp}
\log{P_s^{ZnS}} = 12.812 - 15873/T - [Fe/H]
\end{equation}

\noindent Both equations \ref{eq:kclsvp} and \ref{eq:znssvp} are determined from themochemical equilibrium calculations that depend on published compilations of thermodynamic data for gases and solids, in particular the JANAF tables \citep{fegley1994}.

The different pathways that KCl and ZnS take to transform from gas phase to condensed phase may lead to differences in how KCl and ZnS clouds form. As KCl condensation is a direct phase change, it is amenable to the application of cloud microphysics, where cloud formation begins with nucleation. Homogeneous nucleation occurs when a sufficient number of condensate vapor molecules spontaneously cluster together to surpass the surface energy barrier, while heterogeneous nucleation is similar but occurs on a preexisting surface (condensation nuclei, CNs), which tends to reduce the surface energy barrier \citep{pruppacher1978}. Within the atmospheres of Solar System planets, nucleation sites are typically plentiful enough such that heterogeneous nucleation is the preferred path of cloud formation \citep[e.g.][]{toon1989,moses1992,james1997,colaprete1999,lavvas2011}, though in some cases, such as when CNs are lacking and/or the free energy barrier is low, homogeneous nucleation may occur \citep[e.g.][]{hegg1990,vehkamaki2002,mills2005,friedson2002,shalygina2008}. For exoplanets, the lack of information regarding the nature and distribution of CNs is a challenge for the evaluation of the heterogeneous nucleation rate. Therefore, we consider homogeneous nucleation for KCl cloud formation. The homogeneous nucleation rate is  

\begin{equation}\label{eq:homnuc}
J_{hom} = 4\pi a_c^2 \Phi Z n \exp{(-F/kT)},
\end{equation}

\noindent where $n$ is the number density of condensate vapor molecules, $k$ is the Boltzmann constant, $Z$ is the Zeldovich factor, which takes into account non-equilibrium effects, $\Phi$ is the diffusion rate of vapor molecules to the forming particle, and $F$ is the energy of formation of a particle with radius $a_c$, given by

\begin{equation}
F = \frac{4}{3}\pi \sigma_s a_c^2.
\end{equation}

\noindent where $a_c$ is the critical cluster radius defined as 

\begin{equation}
a_c = \frac{2M\sigma_s}{\rho_p R T \ln{S}},
\end{equation}

\noindent where $M$, $\sigma_s$, $\rho_p$, and $S$ are the molecular weight, surface tension (energy), mass density, and saturation ratio of the condensate, respectively, and $R$ is the universal gas constant. The surface tension of molten KCl is \citep{janz1969}

\begin{equation}
\label{eq:kclsurftens}
\sigma_s^{KCl}(ergs\, cm^{-2}) = 160.4 - 0.07T
\end{equation}

\noindent This value is similar to that found for the \{100\} face of solid KCl crystals \citep{westwood1963}, which is reassuring since it is unknown what phase KCl would be in exoplanet atmospheres. Uncertainties in $\sigma_s^{KCl}$ reported by \citet{janz1969} is $\pm$0.5\%, which is considerably smaller than the factors of 0.5--2 necessary to effect significant variations in cloud distributions \citep{gao2018}.

\begin{figure}[hbt!]
\centering
\includegraphics[width=0.6 \textwidth]{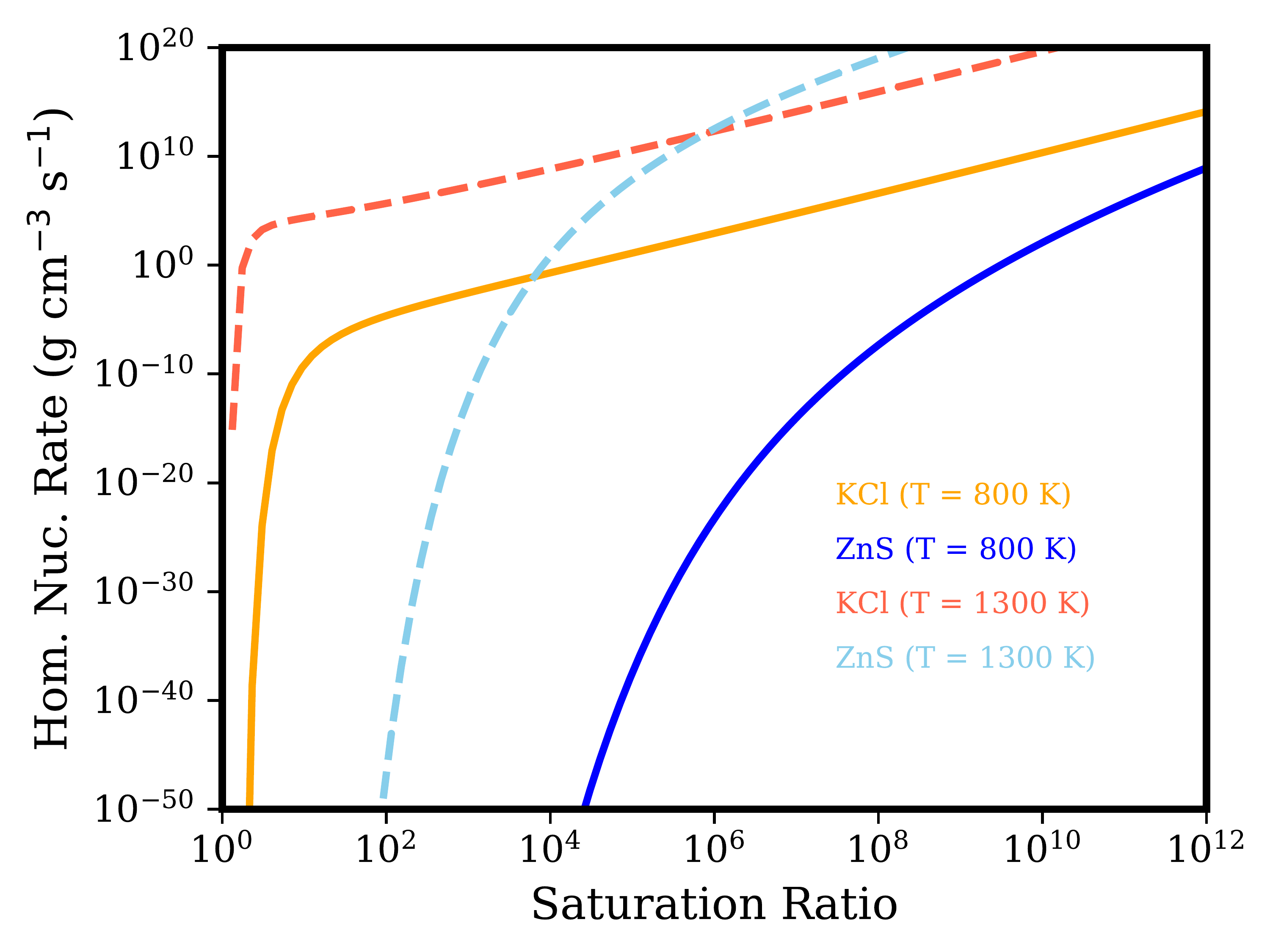}
\caption{The homogeneous nucleation rates of KCl (red and orange) and ZnS (dark and light blue) at temperatures of 800 (solid) and 1300 (dashed) K, expressed as the rate of production of condensed mass within a unit volume, as a function of the saturation ratio of the condensate vapor, defined as the ratio of the partial pressure of the vapor to the saturation vapor pressure of the condensate. We assume $\sigma_s^{ZnS}$ = 860 ergs cm$^{-2}$.}
\label{fig:homnucrate}
\end{figure}

The surface energy data for ZnS is much more varied. \citet{celikkaya1990} estimated a surface energy for ZnS solid, $\sigma_s^{ZnS}$, of 1672 ergs cm$^{-2}$ based on energy of formation arguments; \citet{wright1998} simulated the sphalerite form of ZnS and derived a $\sigma_s^{ZnS}$ value of 650 ergs cm$^{-2}$ for the \{110\} surface, which has the lowest energy; \citet{zhang2003} reports an average surface energy for the sphalerite form of ZnS of 860 ergs cm$^{-2}$ and the wurtzite form of ZnS of 570 ergs cm$^{-2}$, both at 300 K, as calculated by molecular dynamics simulations. They also reviewed a host of previous works that report $\sigma_s^{ZnS}$ values ranging from 900 to 8000 ergs cm$^{-2}$. Although this list is not exhaustive, it is clear that $\sigma_s^{ZnS}$ is much larger than $\sigma_s^{KCl}$, and as such would homogeneously nucleate much slower than KCl for the same temperature, if it is capable of homogeneously nucleating at all \citep{gao2018}. For our work we will set $\sigma_s^{ZnS}$ to a constant 860 ergs cm$^{-2}$, as ZnS is more likely to be in the sphalerite form of ZnS than the wurtzite form given GJ 1214b's atmospheric pressures and temperatures, and the variations of $\sigma_s^{ZnS}$ with temperature are small \citep{zhang2003}. Also, the combined simulation, thermodynamic analysis, and experimental approach of \citet{zhang2003} lends credence to their values. Figure \ref{fig:homnucrate} shows the homogeneous nucleation rates for KCl and ZnS using $\sigma_s^{ZnS}$ = 860 ergs cm$^{-2}$, assuming ZnS can nucleate homogeneously. Even at 1300 K, the highest temperature where ZnS could nucleate, at the ZnS cloud base in the 1000 $\times$ solar metallicity case (Figure \ref{fig:gj1214btp}), the homogeneous nucleation rate is still $\sim$60 of orders of magnitude lower than that of KCl for saturation ratios of 100, which is reached at 10 bars. Therefore, even if ZnS were capable of homogeneous nucleation, the mass of clouds that would form would be dwarfed by that of KCl.

As condensed ZnS forms chemically rather than by direct phase change, it may be more appropriate to model it using grain chemistry. Within the cloud microphysics framework, this can be approximated using heterogeneous nucleation. In other words, ZnS can form by Zn diffusing onto a CN's surface and reacting with H$_2$S there. The heterogeneous nucleation rate in units of critical germs per CN is given by

\begin{equation}\label{eq:hetnuceq}
J_{het} = 4\pi^2 r_{CN}^2 a_c^2 \Phi c_{surf} Z \exp{(-Ff/kT)},
\end{equation}

\noindent where r$_{CN}$ is the radius of the CNs, $c_{surf}$ is the number density of condensate molecules on the nucleating surface, and $f$ is a shape factor related to the ratio of r$_{CN}$ to a$_c$ and the cosine of the contact angle--the angle formed by the CNs' surface and the surface of the critical cluster at the point of contact. The contact angle is defined as the ratio of the difference between the surface energy of the CNs and the interfacial energy of the CNs with the condensate, to the surface energy of the condensate. For context, contact angles for water on foreign surfaces range from $\sim$0$^{\circ}$ for surfaces that are highly hydrophilic, resulting in high heterogeneous nucleation rates, and $>$100$^{\circ}$ for those that are highly hydrophobic, resulting in low heterogeneous nucleation rates \citep{ethington1990,yuan2013}. To approximate grain chemistry, where the rate limiting step is the diffusion and adsorption of condensate molecules to the CN surface \citep{helling2006}, we set the contact angle to a very small value (0.1$^{\circ}$). We do not consider a contact angle of 0$^{\circ}$, as that would indicate no material differences between the CN and condensate.

For the CNs we choose homogeneously nucleated KCl particles. This choice minimizes assumptions of CN sizes and number densities, though ZnS may also condense on other aerosol particles present in the atmosphere, such as photochemical hazes \citep{morley2015,kawashima2018}. KCl was highlighted as a possible CN candidate by \citet{lee2018} due to its relatively high homogeneous nucleation rate. Interactions between the condensate and the CN is defined by, in addition to the contact angle, the desorption energy $F_{des}$ and oscillation frequency $\nu$ of condensate molecules on the CN surface. These quantities are used to define $c_{surf}$,

\begin{equation}
c_{surf} = \frac{\Phi}{\nu}\exp{(F_{des}/kT)}
\end{equation}

\noindent Neither of these quantities are known for KCl and ZnS, and would require laboratory investigations to measure. Large desorption energies ($>$1 eV) are associated with formation of chemical bonds between the adsorbed molecule and the substrate, while small desorption energies ($\sim$0.1 eV) are associated with weaker van der Waals interactions. Desorption energies for small molecules (e.g. H$_2$O, CO$_2$, CH$_4$, etc.) over silicate grains are typically on the order of 0.1 eV in the interstellar medium \citep{seki1983,suhasaria2015,suhasaria2017}, while energies $>$1 eV are possible for metallic surfaces \citep{blaszczyszyn1995,sendecki1989}. \citet{rodriguez1996} measured Zn desorption energies on thin alumina (Al$_2$O$_3$) films of $\sim$25 kcal/mol, or around 1 eV per Zn atom. Zn desorption energies on KCl may be smaller, as it is unlikely to chemically react with KCl. Therefore, a value closer to 0.1 eV is also plausible. The oscillation frequency in units of Hz can be calculated from $F_{des}$ by

\begin{equation}\label{eq:osfreq}
\nu = 1.6 \times 10^{11} \sqrt{F_{des}/M}
\end{equation}

\noindent where $F_{des}$ is in units of temperature, calculated by dividing $F_{des}$ in units of energy by the Boltzmann constant \citep{powell2017}, and $M$ is the molecular weight of Zn. Equation \ref{eq:osfreq} gives an oscillation frequency of 2.13 $\times$ 10$^{12}$ Hz given a desorption energy of 1 eV, and 6.74 $\times$ 10$^{11}$ Hz given a desorption energy of 0.1 eV. Once ZnS has nucleated onto the KCl particle, it is assumed to completely cover it, thereby cutting off KCl vapor from condensing onto the KCl “core”. This is different from the grain chemistry approach of \citep{helling2006} and similar models, where multiple condensates can condense on a particle simultaneously. 

The uncertainties in how ZnS may heterogeneous nucleate on KCl complicates our investigation of how atmospheric and planetary parameters affect cloud distributions on GJ 1214b. Therefore, for comparisons to the observations we will focus solely on pure, homogeneously nucleated KCl clouds. We will investigate mixed ZnS/KCl clouds separately in ${\S}$\ref{sec:mixedclouds}.

\begin{deluxetable}{clcc}
\tablecolumns{4}
\tablecaption{Microphysical parameters \label{table:microparam}}
\tablehead{
\colhead{Symbol} & \colhead{Parameter} & \colhead{Value} & \colhead{References}}
\startdata
\multicolumn{4}{c}{KCl} \\
\hline
$\rho_p$ & Condensed density (g cm$^{-3}$)	&	1.98			&	\nodata		\\
$M$ & Molecular weight (g mol$^{-1}$)	&	74.55			&	\nodata		\\
$P_s^{KCl}$ & Saturation vapor pressure\tablenotemark{a} (dynes cm$^{-2}$)		&	87.3			&	\citet{morley2012}			\\
$\sigma_s^{KCl}$ & Surface energy\tablenotemark{b} (ergs cm$^{-2}$)	&	92			&	\citet{janz1969}		\\
$L$ & Latent heat  (10$^{10}$ ergs g$^{-1}$)			&	2.923			&	\citet{charnay2015b}			\\
$D$ & Molecular diffusion coefficient\tablenotemark{c} (cm$^{2}$ s$^{-1}$)	&	3.96	&		\citet{jacobson2005,sanderson1976}			\\
\nodata & Vapor mixing ratio at depth\tablenotemark{d} (ppmv)	&	0.22			&	\citet{lodders2010}		\\
$d$ & Collision diameter (\AA)	&	2.67		&	\citet{sanderson1976}		\\
\hline
\multicolumn{4}{c}{ZnS} \\
\hline
$\rho_p$ & Condensed density (g cm$^{-3}$)		&	4.09		&\nodata				\\
$M$ & Molecular weight (g mol$^{-1}$)		&	97.474			&	\nodata			\\
$P_s^{ZnS}$ & Saturation vapor pressure\tablenotemark{e} (dynes cm$^{-2}$)		&	344.7		&	\citet{morley2012}		\\
$\sigma_s^{ZnS}$ & Surface energy	(ergs cm$^{-2}$)	&	860		&	\citet{gilbert2003}			\\
$L$ & Latent heat (10$^{10}$ ergs g$^{-1}$)			&	3.118			&	\citet{charnay2015b}			\\
$D$ & Molecular diffusion coefficient\tablenotemark{c} (cm$^{2}$ s$^{-1}$)	&		6.76		&	\citet{jacobson2005,zack2009}			\\
\nodata & Vapor mixing ratio at depth\tablenotemark{d} (ppmv)	&	0.076		&	\citet{lodders2010}		\\
$d$ & Collision diameter (\AA)	&	2.0464		&	\citet{zack2009}		\\
\hline
\multicolumn{4}{c}{Mixed Clouds} \\
\hline
\nodata & Contact angle ($^{\circ}$)	&	0.1		& \nodata	\\
$F_{des}$ & Desorption energy (eV)	&	1, 0.1		& \citet{sendecki1989}	\\
$\nu$ & Oscillation frequency (10$^{12}$ Hz)	&	2.13, 0.674		& \nodata	\\
\enddata
\tablenotetext{a}{At 1 bar and solar metallicity. See Eq. \ref{eq:kclsvp} for variations with temperature.}
\tablenotetext{b}{At 1 bar and solar metallicity. See Eq. \ref{eq:kclsurftens} for variations with temperature.}
\tablenotetext{c}{At 1 bar and solar metallicity. See Eq. \ref{eq:diffus} for variations with temperature.}
\tablenotetext{d}{At solar metallicity.}
\tablenotetext{e}{At 1 bar and solar metallicity. See Eq. \ref{eq:znssvp} for variations with temperature and metallicity.}
\end{deluxetable}

Several additional material properties are needed to define a condensate within CARMA. The full list of properties and their values are given in Table \ref{table:microparam}. The latent heat of vaporization $L$ of KCl and ZnS is calculated from their saturation vapor pressures and the Clausius--Clapeyron equation

\begin{equation}
\label{eq:latentheat}
\frac{d\ln{P_s^x}}{dT} = \frac{L}{RT^2}
\end{equation}

\noindent \citep{charnay2015b}. For ZnS, the latent heat of vaporization can be considered energy released at the formation of ZnS from Zn and H$_2$S.

Condensation of trace gases in an atmosphere depends strongly on the diffusion rate of gas molecules to the surface of the cloud particles. The molecular diffusion coefficient $D$ can be expressed using Chapman--Enskog theory as \citep{jacobson2005}

\begin{equation}
\label{eq:diffus}
D = \frac{5}{16 A d^2 \rho_a \Omega_c} \sqrt{\frac{R T M_a (M + M_a)}{2 \pi M}}
\end{equation}

\noindent where $A$ is Avogadro's number, $\rho_a$ is the atmospheric mass density, $M_a$ is the mean molecular weight of the atmosphere, $\Omega_c$ is the collision integral of the condensing species with the main atmospheric components and is of order unity, and $d$ is the collision diameter of the condensing species. We estimate the collision diameters for KCl and ZnS to be 2.67 $\AA$ and 2.0464 $\AA$, respectively, from their equilibrium bond lengths \citep{sanderson1976,zack2009}. 
Abundances of the condensates, KCl and ZnS, are assumed to correspond to those of K and Zn at the designated metallicities, as they are both the limiting element in their molecules. The volume mixing ratios of K and Zn in a solar metallicity atmosphere are 0.22 and 0.076 ppmv, respectively \citep{lodders2010}. We assume their abundances scale linearly with metallicity.  

\subsection{Model Setup}

Each model atmosphere is discretized into 40 layers from 10 kbar to 0.1 nbar equidistant in log--pressure space. The cloud particle size distribution is divided into 65 bins, starting from 1 \AA particles and doubling in mass with each subsequent bin.

All simulations are initiated in an atmosphere devoid of particles and vapor, except the deepest atmospheric pressure level, which possesses fixed condensate volume mixing ratios corresponding to their solar abundances modulated by the metallicity as a bottom boundary condition. The particle number density is fixed to zero at the bottom boundary due to the high temperatures there. Upon initiating the simulation, condensate vapor is mixed upwards by eddy diffusion until it reaches saturation, at which point the vapor can nucleate to produce cloud particles. For the pure KCl cloud simulations, KCl is allowed to homogeneously nucleate. For the mixed cloud simulations, KCl homogeneously nucleates while ZnS heterogeneously nucleate onto the pure KCl particles. 

Following nucleation, the cloud particles can grow by condensation or be lost through evaporation as the overall cloud distribution evolves towards equilibrium. The cloud particles are also transported via eddy diffusion and sedimentation under gravity. We do not consider coagulation in our simulations, as it is uncertain how these cloud particles would be transformed by collisions. We discuss coagulation further in ${\S}$\ref{sec:coalcon}. We set a zero--flux top boundary condition, corresponding to a lack of significant loss of condensing/condensed material to the upper atmosphere.  

\section{Results}\label{sec:results5}

\subsection{Oscillatory Behavior}

\begin{figure}[hbt!]
\centering
\includegraphics[width=0.6 \textwidth]{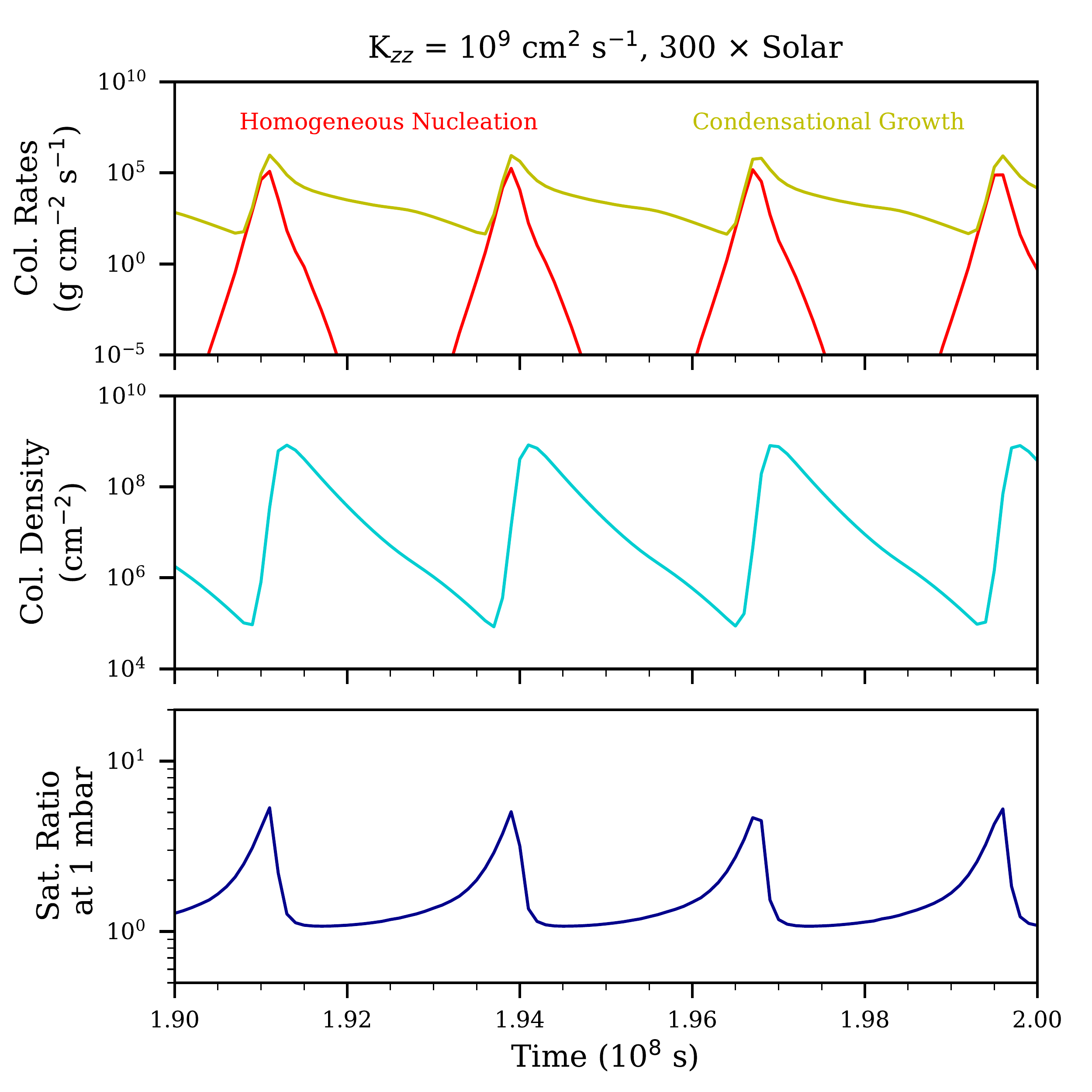}
\caption{Oscillations in the KCl cloud distribution as a function of time, as shown by the column--integrated homogeneous nucleation and condensational growth rates (red and yellow, respectively, top) and particle number density (middle), and the saturation ratio of KCl vapor at 1 mbar (bottom). A \kzz value of 10$^{9}$ cm$^{2}$ s$^{-1}$ and a metallicity of 300 $\times$ solar are assumed. }
\label{fig:oscillations}
\end{figure}

Our results show oscillatory behavior in the KCl cloud distribution as a function of time. Figure \ref{fig:oscillations} shows how this oscillation is expressed in the 300 $\times$ solar metallicity, K$_{zz}$ = 10$^{9}$ cm$^{2}$ s$^{-1}$ case. The period of these oscillations are $\sim$1 Earth month. Similar oscillations are seen in all of our pure KCl cloud cases, though their periods and amplitudes differ, with higher \kzz cases having longer periods and lower amplitudes. 

This phenomenon may stem from a real physical process occurring in these clouds, akin to a ``rain cycle'' \citep{barth2003,gao2014,parkinson2015}. As Figure \ref{fig:oscillations} shows, the homogeneous nucleation of KCl particles occurs in pulses that take place at maximum supersaturation; each pulse increases the number of cloud particles and the overall growth rate of particles. The saturation ratio drops precipitously after the nucleation pulse due to formation and growth of new particles depleting the condensate vapor; the drop in saturation then shuts down nucleation, leading to a gradual decrease in particle number density and growth as particles sediment out of the cloud; this decrease in vapor uptake by particles leads to a build-up of vapor upwelled from below the cloud via eddy diffusion, and the cycle repeats. 

Whether this cycle actually occurs in exoplanet atmospheres is difficult to assess. Actual clouds are 3--dimensional systems controlled by advection of vapor and particles in various directions instead of 1D eddy diffusion, though advection can also produce local pockets of supersaturation that causes high homogeneous nucleation rates \citep{helling2001}. In order to address this issue a coupled dynamics--microphysical model will be needed, which is beyond the scope of this work. Therefore, for the remainder of this paper we will show time--averaged cloud and vapor distributions. 

\subsection{Timescales}\label{sec:timescales}

\begin{figure}[hbt!]
\centering
\includegraphics[width=0.8 \textwidth]{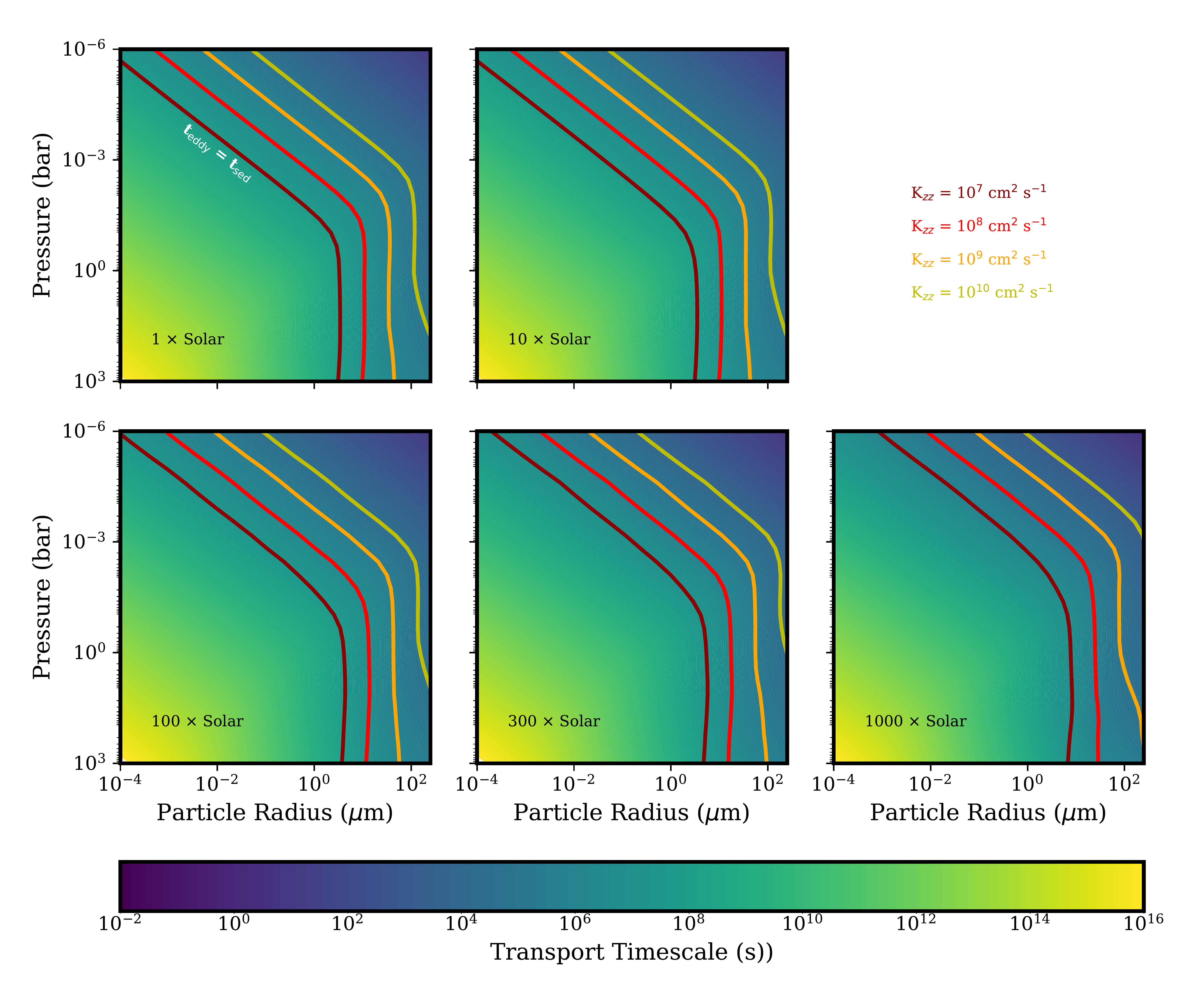}
\caption{Timescales of transport by sedimentation and eddy diffusion as a function of pressure level in the atmosphere and cloud particle radius, for atmospheres with 1 $\times$ (top left), 10 $\times$ (top middle), 100 $\times$ (bottom left), 300 $\times$ (bottom middle), and 1000 $\times$ (bottom right) solar metallicity. The colors show the sedimentation timescale, while the curves indicate where the sedimentation timescale equals the eddy diffusion timescale for K$_{zz}$ = 10$^{7}$ (dark red), 10$^{8}$ (red), 10$^{9}$ (orange), and 10$^{10}$ (yellow) cm$^{2}$ s$^{-1}$. Transport is dominated by eddy diffusion to the left of the curves, and by sedimentation to the right of the curves.}
\label{fig:transtime}
\end{figure}

\begin{figure}[hbt!]
\centering
\includegraphics[width=1.0 \textwidth]{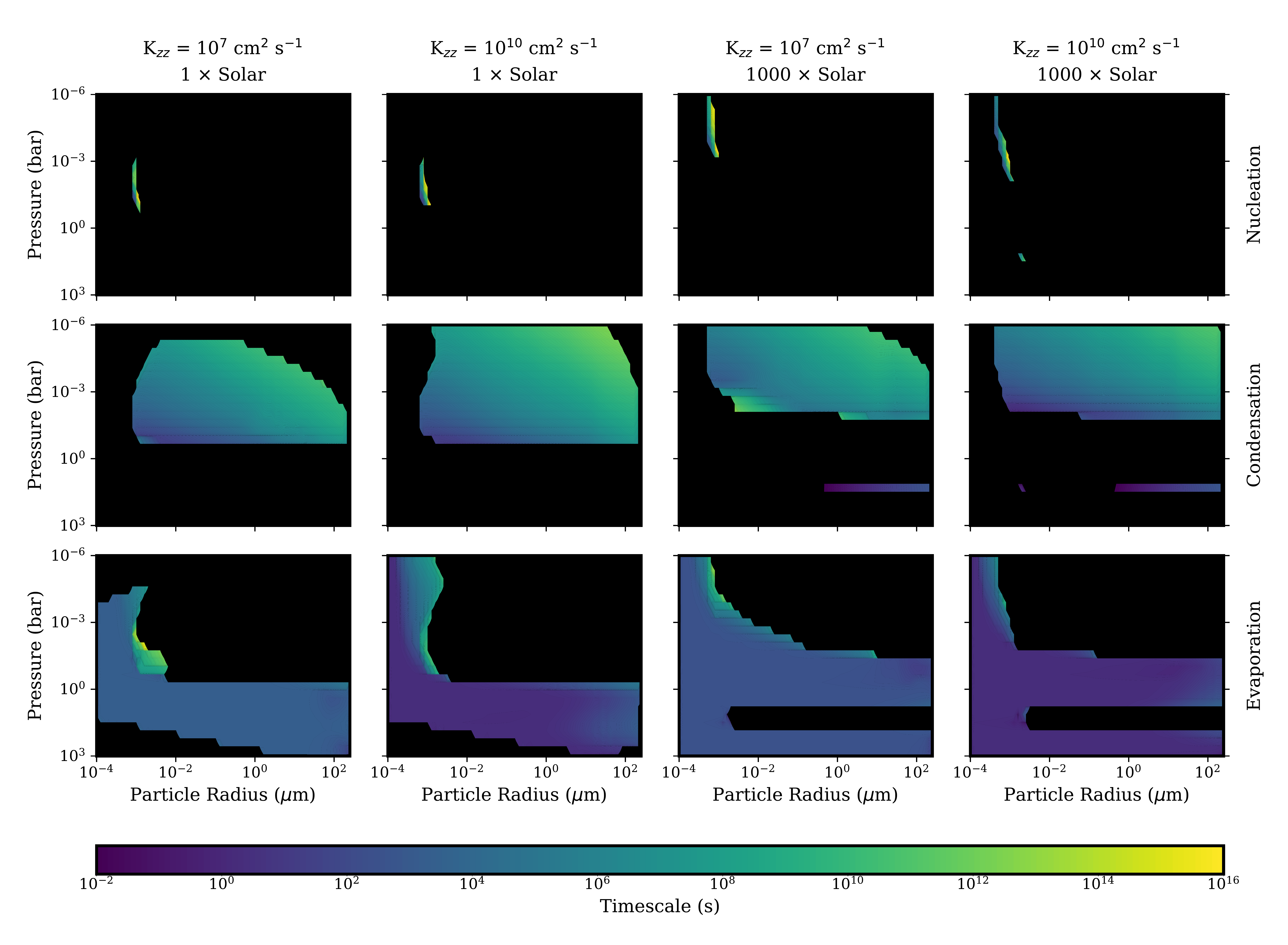}
\caption{Timescales of nucleation (top), growth by condensation (middle), and evaporation (bottom) for the K$_{zz}$ = 10$^{7}$ cm$^{2}$ s$^{-1}$ and 1 $\times$ solar metallicity (left), K$_{zz}$ = 10$^{10}$ cm$^{2}$ s$^{-1}$ and 1 $\times$ solar metallicity (left middle); K$_{zz}$ = 10$^{7}$ cm$^{2}$ s$^{-1}$, 1000 $\times$ solar metallicity (right middle), and K$_{zz}$ = 10$^{10}$ cm$^{2}$ s$^{-1}$, 1000 $\times$ solar metallicity (right) cases.}
\label{fig:microtime}
\end{figure}

The timescales associated with nucleation, condensation, evaporation, and transport give insight into the processes sculpting the equilibrium cloud distribution. Figures \ref{fig:transtime} and \ref{fig:microtime} show these timescales. The sedimentation timescale $t_{sed}$ is defined as $H/v_f$, where $H$ is the scale height and $v_f$ is the particle fall velocity. The eddy diffusion timescale $t_{eddy}$ is defined as $H^2/K_{zz}$. As expected, $t_{sed}$ increases deeper in the atmosphere and for smaller particles due to the dependence of $v_f$ on atmospheric viscosity and the square of the particle radius. As $t_{eddy}$ depends on $H$, which does not change significantly throughout the atmosphere compared to $\rho_a$, and \kzz has been assumed to be constant, $t_{eddy}$ is nearly constant compared to $t_{sed}$. The curves in Figure \ref{fig:transtime} show where $t_{sed}$ = $t_{eddy}$ for the four \kzz cases. Regions of phase space to the left of the curves (larger particles and lower pressure levels) are dominated by sedimentation ($t_{sed}$ $<$ $t_{eddy}$), while regions of phase space to the right of the curves (smaller particles and higher pressure levels) are dominated by eddy diffusion. The $t_{sed}$ = $t_{eddy}$ curves move towards lower pressure levels and larger particles with increasing metallicity due to decreasing $H$ reducing $t_{eddy}$. A more physical treatment of \kzz where it is proportional to $H$ (e.g. mixing length theory) may reduce this effect. Finally, a combined transport timescale $t_{trans}$ can be defined as the minimum value between $t_{sed}$ and $t_{eddy}$ at each pressure level and particle radius bin. 

The nucleation, condensation, and evaporation timescales $t_{nuc}$, $t_{cond}$, and $t_{evap}$, respectively, are all defined according to $m_p/(dm_p/dt)$, where $m_p$ is the mass density of particles in a mass bin, and $dm_p/dt$ is the rate of change in particle mass density in said bin due to the three separate microphysical processes, as calculated by CARMA. For (homogeneous) nucleation, $dm_p/dt$ = $J_{hom}$ (Equation \ref{eq:homnuc}), while for condensation and evaporation we refer the reader to Equation 34 in \citet{gao2018}.

The nucleation timescale is only relevant for a tiny region of phase space, specifically for radii $\sim a_c$.  The vertical extent of the nucleation region is defined by $S$ and $T$ such that $J_{hom}$ is maximized. At lower temperatures (lower pressure levels), though $S$ may be larger, the exponential term $F/kT$ becomes large as well, reducing $J_{hom}$. At higher temperatures (higher pressure levels), $S$ decreases, again reducing $J_{hom}$. Higher metallicities result in upward movement of the nucleation region (and thus cloud base) in the atmosphere due to increased $T$. As alluded to in ${\S}$\ref{sec:modeldes}, KCl also nucleates at $\sim$30 bars in the 1000 $\times$ solar metallicity case due to the condensation curve of KCl crossing the PT pressure twice. Higher \kzz results in more extended regions of short nucleation timescale due to increased supply of condensate vapor from depth, which maintains a high $S$.

The condensation timescale is finite for a large part of phase space and tends to increase with increasing particle size. This is related to the decreasing surface area to volume ratio of particles as they become larger, and thus the fractional change in particle mass decreases as the mass of the particle increases. The condensation timescale also increases with height in the atmosphere as the abundance of condensate vapor decreases at lower pressure levels. Increasing \kzz results in more condensate vapor at lower pressure levels, reducing the condensation timescale there. Particle growth at depth for the 1000 $\times$ solar metallicity case is extremely rapid due to large abundances of condensate vapor.

The evaporation timescale is finite for much of the phase space not dominated by condensation, and is in general much shorter than the condensation and nucleation timescales. Evaporation occurs for all particles smaller than $a_c$ due to instabilities caused by the Kelvin curvature effect, and is also fast for particles below the cloud deck due to high temperatures. The evaporation timescale is shorter for high \kzz due to the higher abundance of smaller particles, which evaporate faster. The smaller average particle size is caused by higher nucleation rates producing a large number of small particles that fail to grow larger due to the finite condensate vapor supply \citep{gao2018}.

Higher temperatures resulting from higher metallicities move the condensation--evaporation boundary higher in the atmosphere. However, there are areas of overlap where particles are both evaporating and growing simultaneously. This is likely a numerical effect resulting from a net change in mass that is close to zero and may oscillate back and forth within a single time step. In such a situation, the process with a shorter timescale should dominate

\subsection{Cloud and Vapor Distributions}\label{sec:kzzmetvar}

\begin{figure}[hbt!]
\centering
\includegraphics[width=1.0 \textwidth]{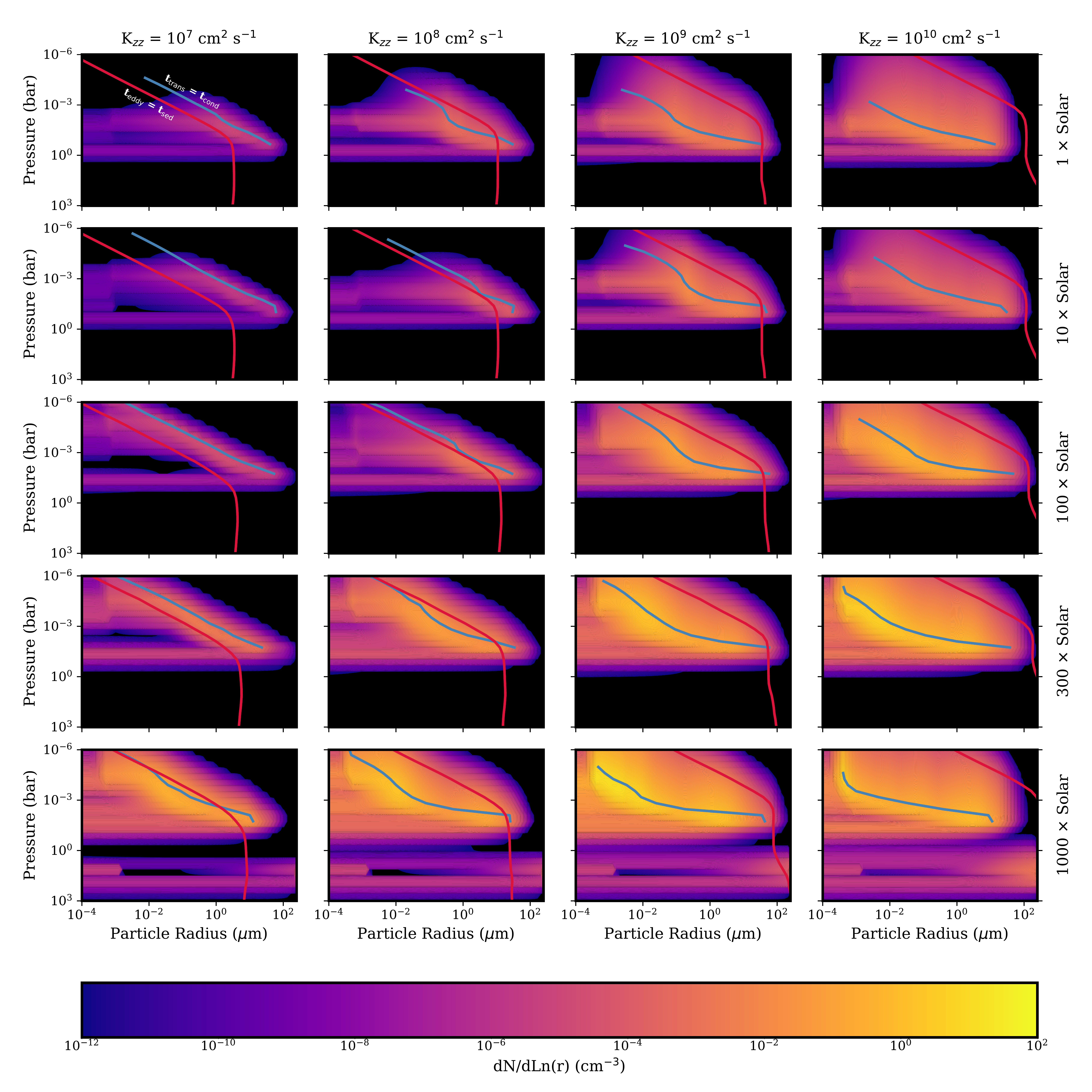}
\caption{KCl cloud particle number density ($dN/dLn(r)$) as a function of particle radius and atmospheric pressure level for all pure KCl cloud cases. The blue curves show where the timescale of growth by condensation equals the transport timescale, while the red curves show where the eddy diffusion timescale equals the sedimentation timescale.}
\label{fig:allcontour}
\end{figure}

The KCl cloud particle distribution morphology differs significantly as \kzz and metallicity are varied, as shown in Figure \ref{fig:allcontour}. These variations can be understood by comparing Figures \ref{fig:transtime} and \ref{fig:microtime} and noting where the various timescales equal each other. In particular, the curve delineating where $t_{cond}$ = $t_{trans}$ (blue curve in Figure \ref{fig:allcontour}) coincides with the peak in particle number density as a function of atmospheric pressure level. This can be understood by considering what happens to particles on either side of this curve. Smaller particles grow faster than they can be transported, and so quickly reach the peak radius. Larger particles grow slower than they can be transported, and so their distribution is controlled by eddy diffusion or sedimentation, which in turn depends on which side of the $t_{eddy}$ = $t_{sed}$ boundary (red curve in Figure \ref{fig:allcontour}) they exist on.

In cases where the $t_{cond}$ = $t_{trans}$ curve is above the $t_{eddy}$ = $t_{sed}$ curve, large particles are transported mainly by sedimentation. As such, very few particles larger than the $t_{cond}$ = $t_{trans}$ curve exist, as they have all rained out. In cases where the $t_{cond}$ = $t_{trans}$ curve is below the $t_{eddy}$ = $t_{sed}$ curve, large particles are transported mainly by eddy diffusion. As such, particles larger than the $t_{cond}$ = $t_{trans}$ curve originate as particles on the $t_{cond}$ = $t_{trans}$ curve, but which have been lofted to lower pressure levels. In particular, these particles are lofted up to the $t_{eddy}$ = $t_{sed}$ curve, above which sedimentation greatly reduce particle number densities. In this way, the $t_{cond}$ = $t_{trans}$ and $t_{eddy}$ = $t_{sed}$ curves help define the extent of the cloud in pressure level--particle radii space.

Evaporation dominates below the cloud deck, as $t_{evap}$ is very short. This is shown by the horizontal ``bar'' that arises at the lower tip of the $t_{cond}$ = $t_{trans}$ curve, and which stretches towards smaller particle radii, showing the transformation of large particles into smaller particles due to evaporation. As such, evaporation at the cloud base prevents larger particles from forming within the cloud. In other words, even when eddy diffusion allows for larger particles to be kept aloft against sedimentation, these particles cannot exist because the maximum cloud particle size is controlled by condensation and evaporation. On the other hand, particles larger than that which can be kept aloft by eddy diffusion can still exist if growth by condensation is fast enough to outpace sedimentation. The vital role of $t_{cond}$ in defining the cloud distribution underlines the importance of considering microphysical processes in addition to balancing sedimentation and eddy diffusion.

The 1000 $\times$ solar metallicity cases exhibit cloud formation below the main cloud deck due to the double crossing of the saturation vapor pressure (Figure \ref{fig:gj1214btp}). The particles in these deep cloud decks are much larger than those in the main cloud decks ($>100$ $\mu$m vs. 10s of $\mu$m) due to increased condensate vapor abundances. Again, these particles can be larger than that allowed by eddy diffusion because of their short $t_{cond}$.

None of the particle size distributions are strictly lognormal, as has been proposed for previous exoplanet cloud models \citep[e.g.][]{ackerman2001}, particularly those cases where the $t_{cond}$ = $t_{trans}$ curve lies below the $t_{eddy}$ = $t_{sed}$ curve. In such cases, the particle size distribution tends to have a peak coinciding with the $t_{cond}$ = $t_{trans}$ curve, followed by a nearly flat ``plateau'' stretching out to the maximum particle size in the cloud as defined by the particle size at the cloud base, and the particle radius at the $t_{eddy}$ = $t_{sed}$ curve. \citet{gao2018} showed that the maximum difference in optical properties between a cloud possessing such a size distribution and a cloud with a lognormal particle size distribution is about a factor of 2, provided the lognormal distribution is peaked at slightly smaller sizes than the maximum particle size on the ``plateau'' of the irregular distribution.

\begin{figure}[hbt!]
\centering
\includegraphics[width=1.0 \textwidth]{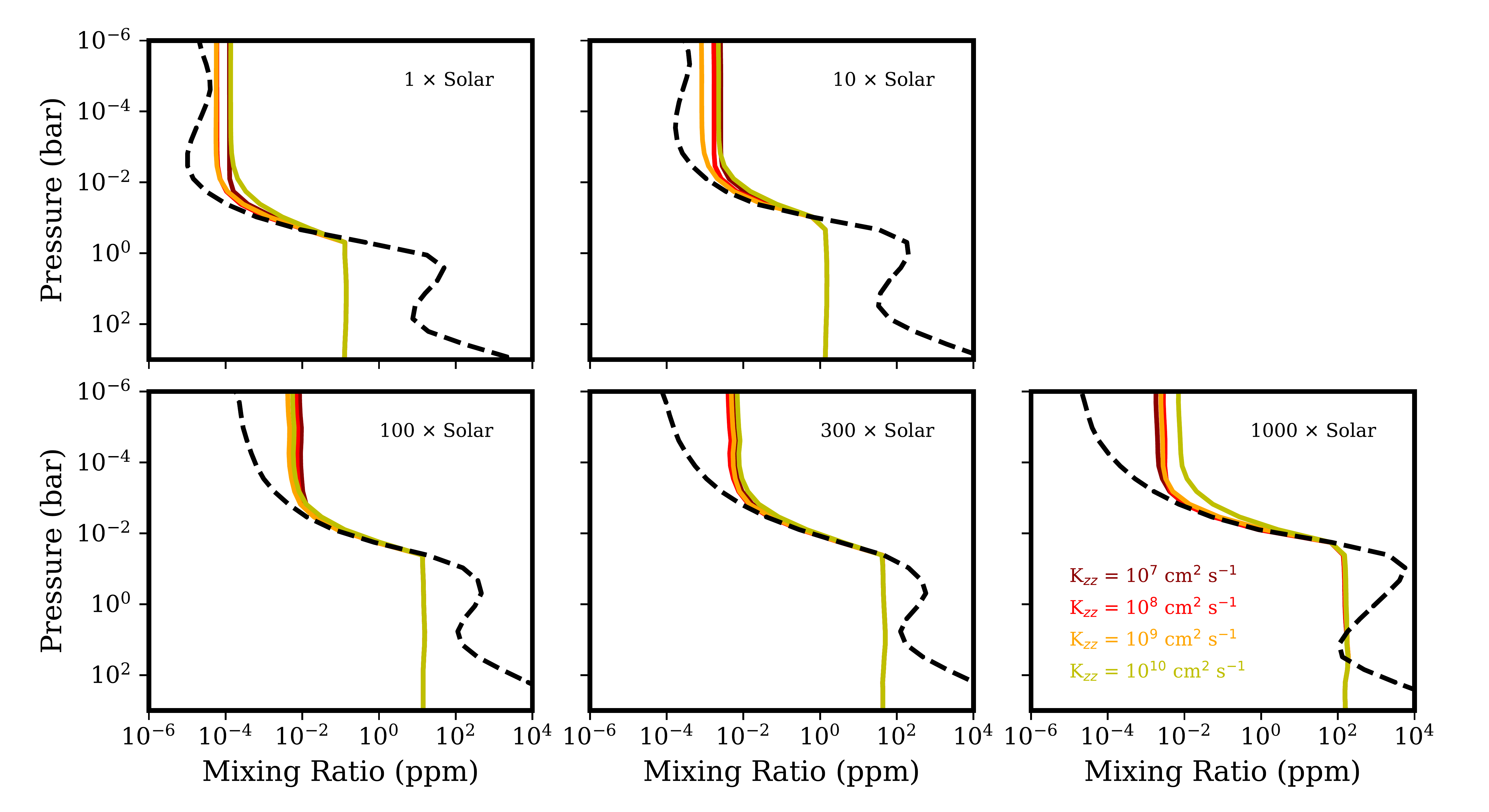}
\caption{KCl vapor mixing ratio for the 1 $\times$ (top left), 10 $\times$ (top middle), 100 $\times$ (bottom left), 300 $\times$ (bottom middle) and 1000 $\times$ (bottom right) solar metallicity cases, and for \kzz values of 10$^{7}$ (dark red), 10$^{8}$ (red), 10$^{9}$ (orange), and 10$^{10}$ (yellow) cm$^{2}$ s$^{-1}$, compared to the KCl saturation vapor mixing ratio (black, dashed).}
\label{fig:metvapor}
\end{figure}

\begin{figure}[hbt!]
\centering
\includegraphics[width=0.6 \textwidth]{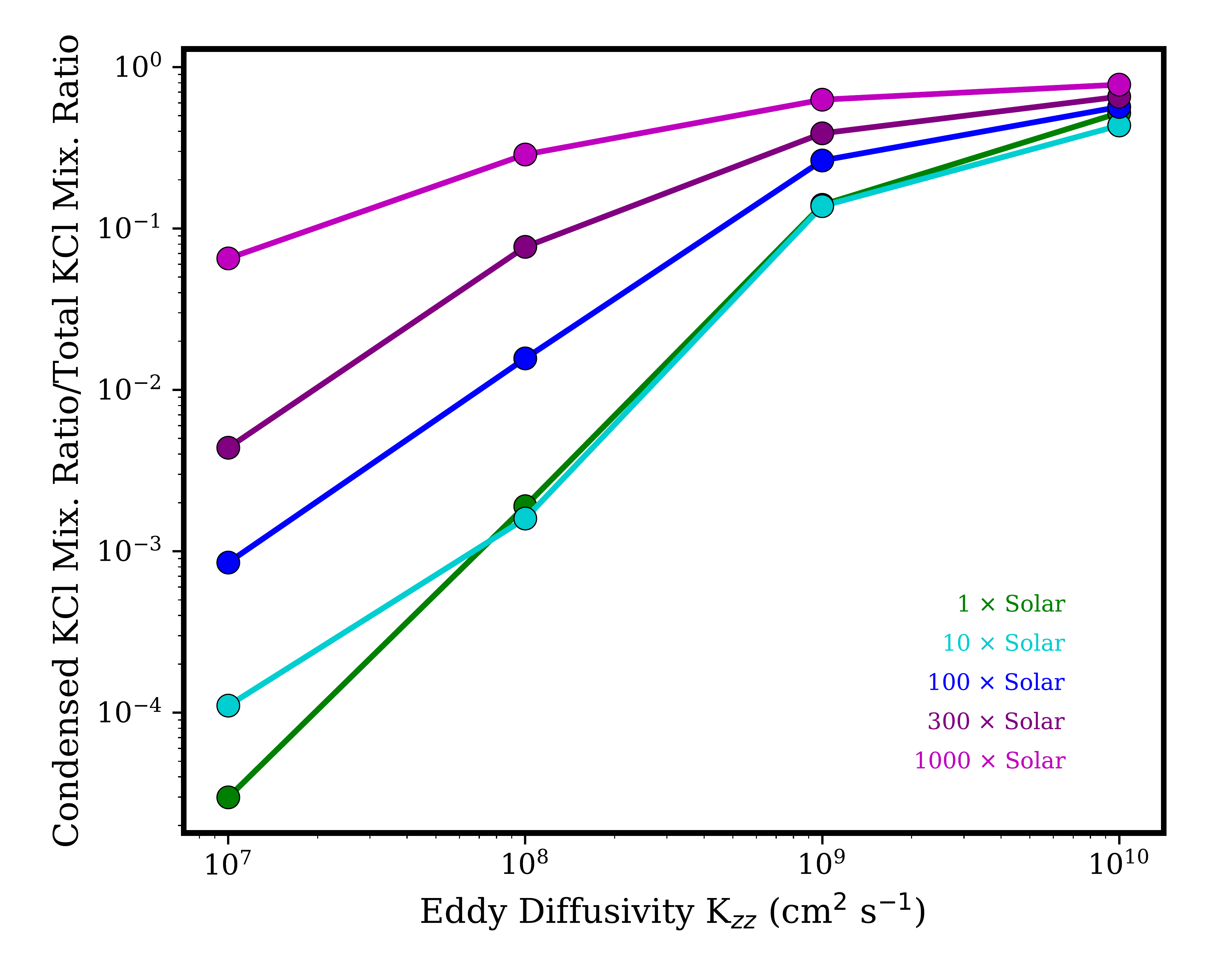}
\caption{Condensed KCl mixing ratio above 10 mbar in units of the total KCl mixing ratio as a function of \kzz for 1 $\times$ (green), 10 $\times$ (cyan), 100 $\times$ (blue), 300 $\times$ (purple), and 1000 $\times$ (magenta) solar metallicity.}
\label{fig:colmass}
\end{figure}

The KCl vapor mixing ratio above the cloud is set by the balance of vapor loss due to cloud formation and upwelling of vapor from depth due to eddy diffusion. Figure \ref{fig:metvapor} presents the KCl vapor mixing ratio for all \kzz and metallicity cases, and shows that, at a given metallicity, the KCl vapor mixing ratio is within an order of magnitude of each other for all \kzz values. As the upward flux of vapor increases with increasing \kzz, it suggests that the loss rate of vapor to cloud formation also increases with increasing \kzz. Indeed, the mass of condensed KCl increases with increasing \kzz for all metallicities (Figure \ref{fig:colmass}). However, this mass increase reaches a limit at high \kzz, when the total KCl abundance (condensed and vapor) becomes nearly fully mixed within the atmosphere. In other words, downward mixing of cloud particles becomes efficient, thereby limiting the cloud mass, even though cloud production (and corresponding vapor loss) continues to rise with \kzz.

\subsection{Cloud Optical Depth}\label{sec:cloudtau}

\begin{figure}[hbt!]
\centering
\includegraphics[width=1.0 \textwidth]{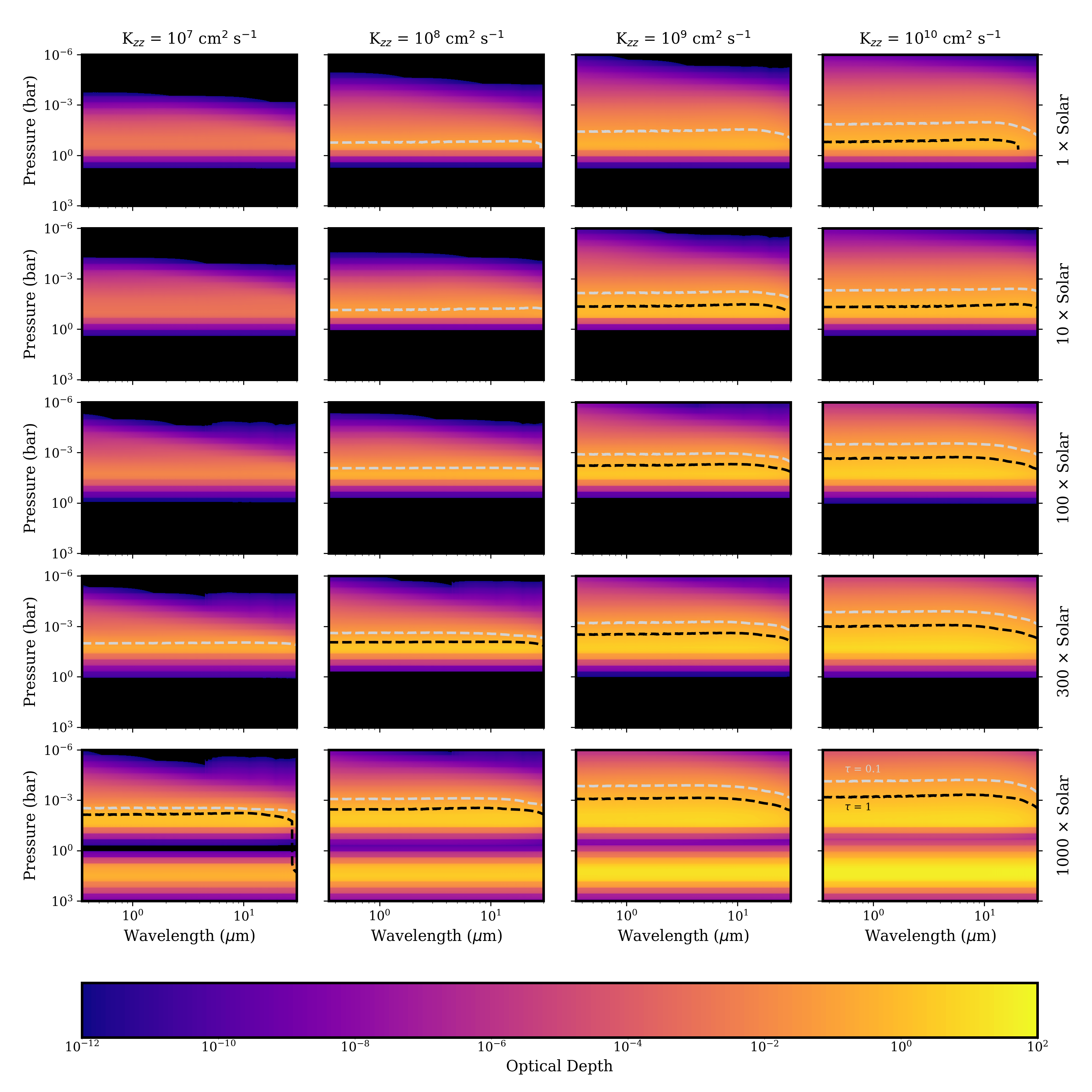}
\caption{KCl cloud optical depth as a function of wavelength and atmospheric pressure level for all pure KCl cases. Grey and black dashed curves show the pressure levels at which the nadir optical depth equals 0.1 and 1, respectively. }
\label{fig:taufig}
\end{figure}

\begin{figure}[hbt!]
\centering
\includegraphics[width=0.8 \textwidth]{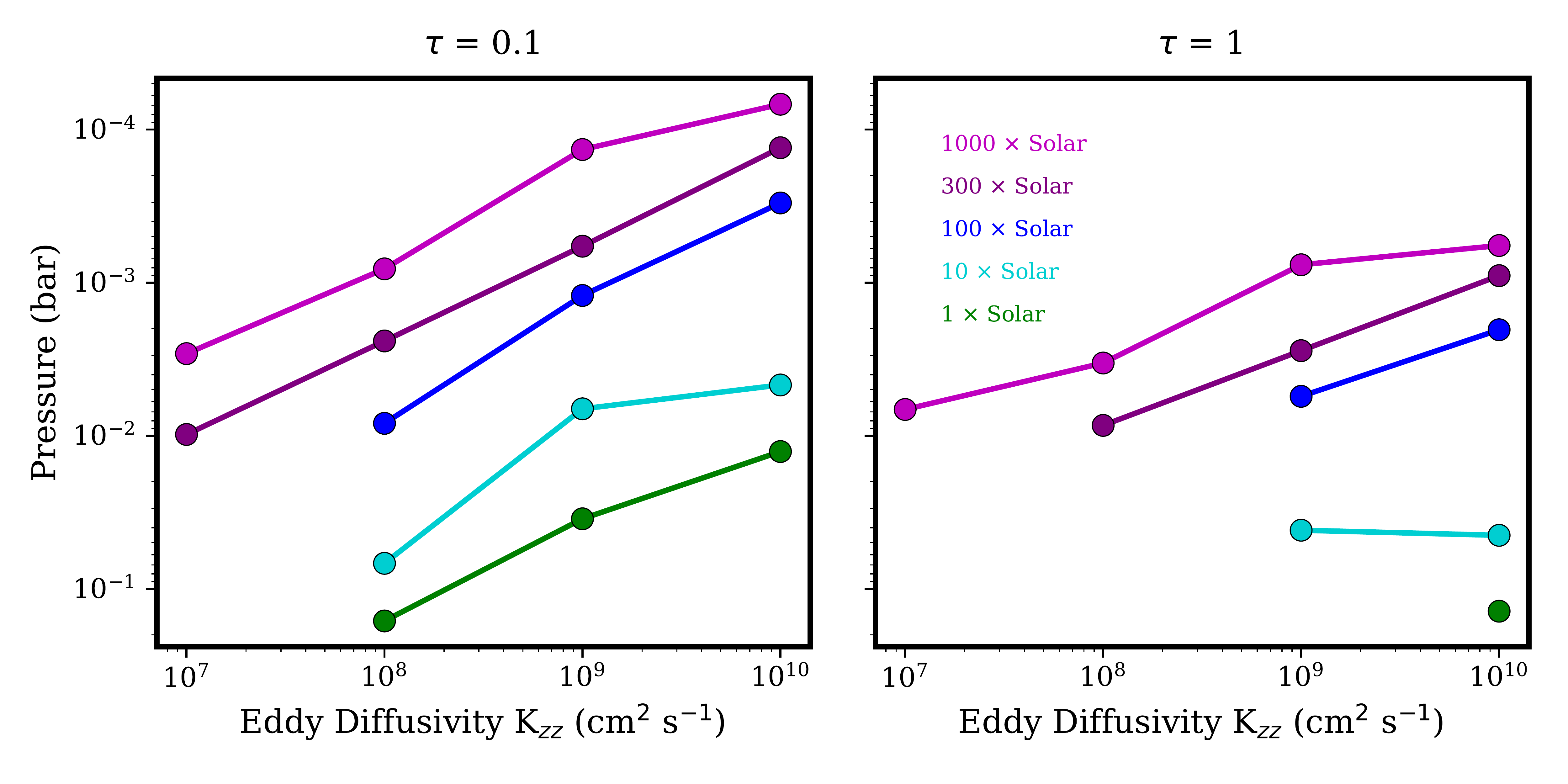}
\caption{Optical depth equals 0.1 (left) and 1 (right) pressure levels as a function of \kzz for 1 $\times$ (green), 10 $\times$ (cyan), 100 $\times$ (blue), 300 $\times$ (purple), and 1000 $\times$ (magenta) solar metallicity.}
\label{fig:taugrid}
\end{figure}

We calculate the optical depth $\tau$ of our simulated pure KCl clouds as a function of atmospheric pressure level and wavelength from 0.35 to 30 $\mu$m assuming Mie theory, as shown in Figure \ref{fig:taufig}. Also shown, in dashed lines, are the pressure levels where $\tau$ = 0.1 and 1, which are presented explicitly in Figure \ref{fig:taugrid} averaged over the wavelength range 1--5 $\mu$m. As expected, the KCl clouds become more opaque as \kzz and metallicity increases, though at high \kzz the $\tau$ increase may be reduced due to reaching the fully mixed limit. In addition, the cloud spectrum is essentially flat with two minor deviations: A slight downward slope towards shorter wavelengths shortward of a few microns caused by the peak in extinction efficiency at 2$\pi$r/$\lambda$ $\sim$ 1 in Mie theory, and a decrease in opacity at wavelengths much longer than the size of the cloud particles. For the 1000 $\times$ metallicity cases an obvious deep opaque cloud is present. These results show no sign of Rayleigh scattering by small cloud particles, as the similar-in-number-density larger particles dominate the opacity \citep{wakeford2015}. Finally, by assuming that the cloud top is between the $\tau$ $\sim$ 0.1--1 pressure levels, we find that the higher \kzz and higher metallicity cases are capable of generating cloud tops consistent with the observations of \citet{kreidberg2014}.

\subsection{Comparison with Observations}\label{sec:compobs}

\begin{figure}[hbt!]
\centering
\includegraphics[width=1.0 \textwidth]{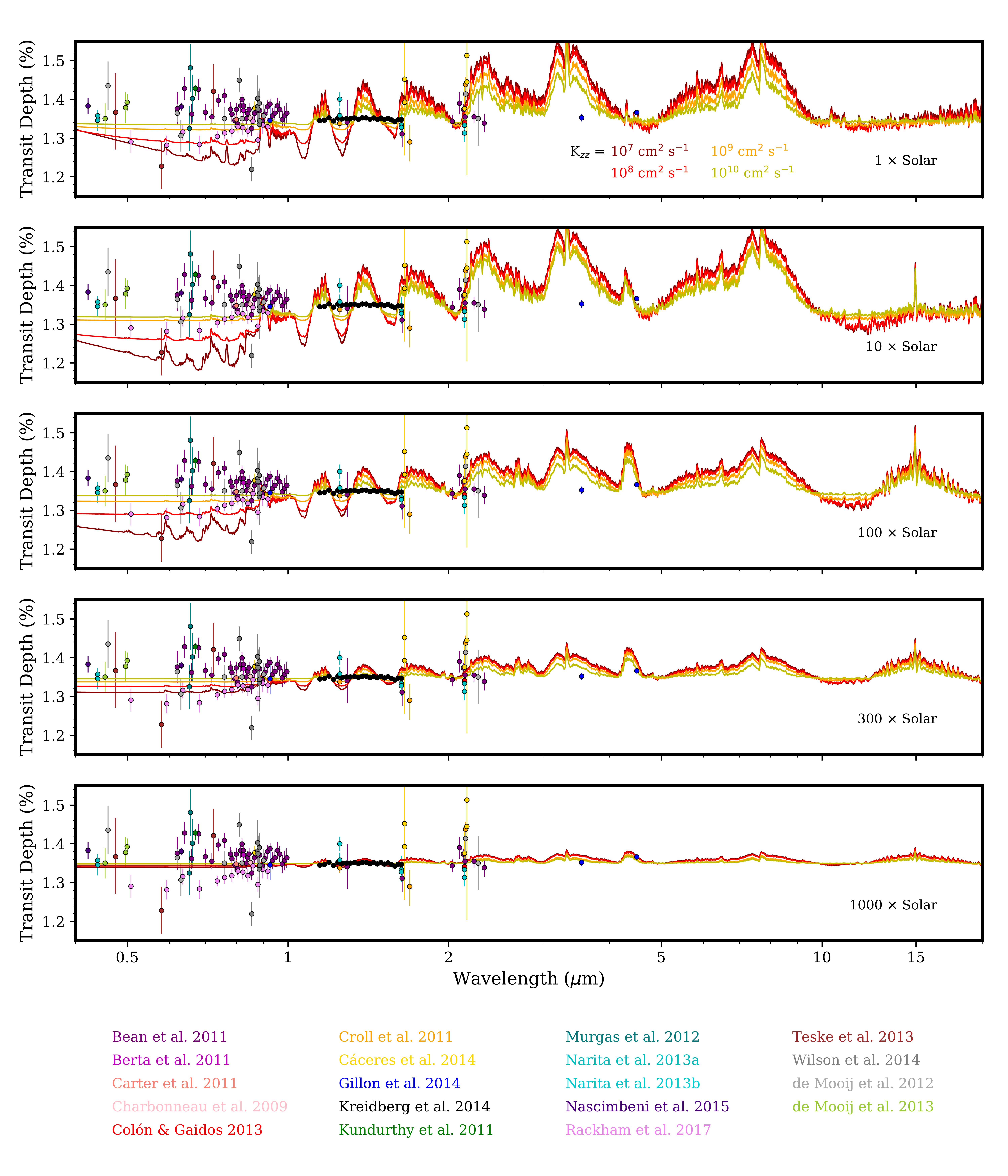}
\caption{Comparison of synthetic transmission spectra generated from our cloudy atmospheres with all available GJ 1214 b transmission spectra observations. Our models are presented as follows: the K$_{zz}$ = 10$^{7}$ (dark red), 10$^{8}$ (red), 10$^{9}$ (orange), and 10$^{10}$ (yellow) cm$^{2}$ s$^{-1}$ cases are shown for metallicities of, listed from the top down, 1 $\times$, 10 $\times$, 100 $\times$, 300 $\times$, and 1000 $\times$ solar.}
\label{fig:compobsall}
\end{figure}

We generate synthetic transmission spectra from our simulated cloud distributions using SCARLET, and fit them to all available observations of GJ 1214 b \citep{charbonneau2009,bean2011,berta2011,carter2011,croll2011,kundurthy2011,murgas2012,demooij2012,colon2013,narita2013a,narita2013b,teske2013,demooij2013,caceres2014,gillon2014,kreidberg2014,wilson2014,nascimbeni2015,rackham2017}. The only tunable parameter in the fit is the base (wavelength--independent) planetary radii, which moves the entire synthetic transmission spectrum to larger or small transit depths by the same amount at all wavelengths. Figure \ref{fig:compobsall} shows our best fits for our cases. 

As can be discerned from Figure \ref{fig:taufig}, the low \kzz cloud cases do not significantly alter the transmission spectrum from that of a clear atmosphere, while higher \kzz clouds cause flattening of spectral features due to increased cloud opacity. Increasing metallicity decreases the magnitude of the spectral features due to increased mean atmospheric molecular weight, as well as enhance the CO and CO$_{2}$ features at 4.5 and 15 microns, respectively.

\begin{figure}[hbt!]
\centering
\includegraphics[width=1.0 \textwidth]{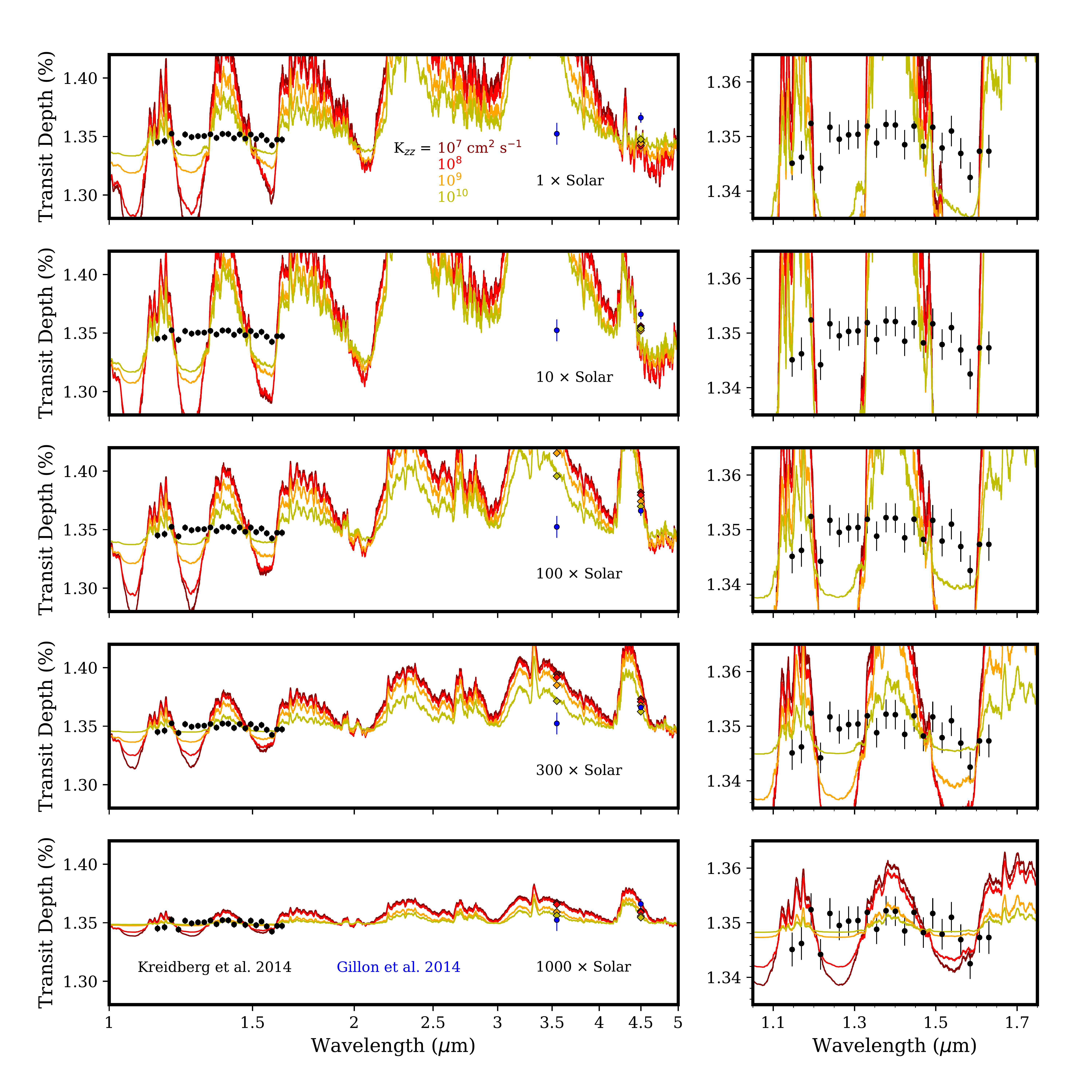}
\caption{Same as Figure \ref{fig:compobsall}, but comparing to only space-based observations (left) and to only WFC3 observations by \citet{kreidberg2014} (right). Note the different x and y axis ranges for the different subplots.}
\label{fig:compobsspacewfc3}
\end{figure}

\begin{figure}[hbt!]
\centering
\includegraphics[width=0.8 \textwidth]{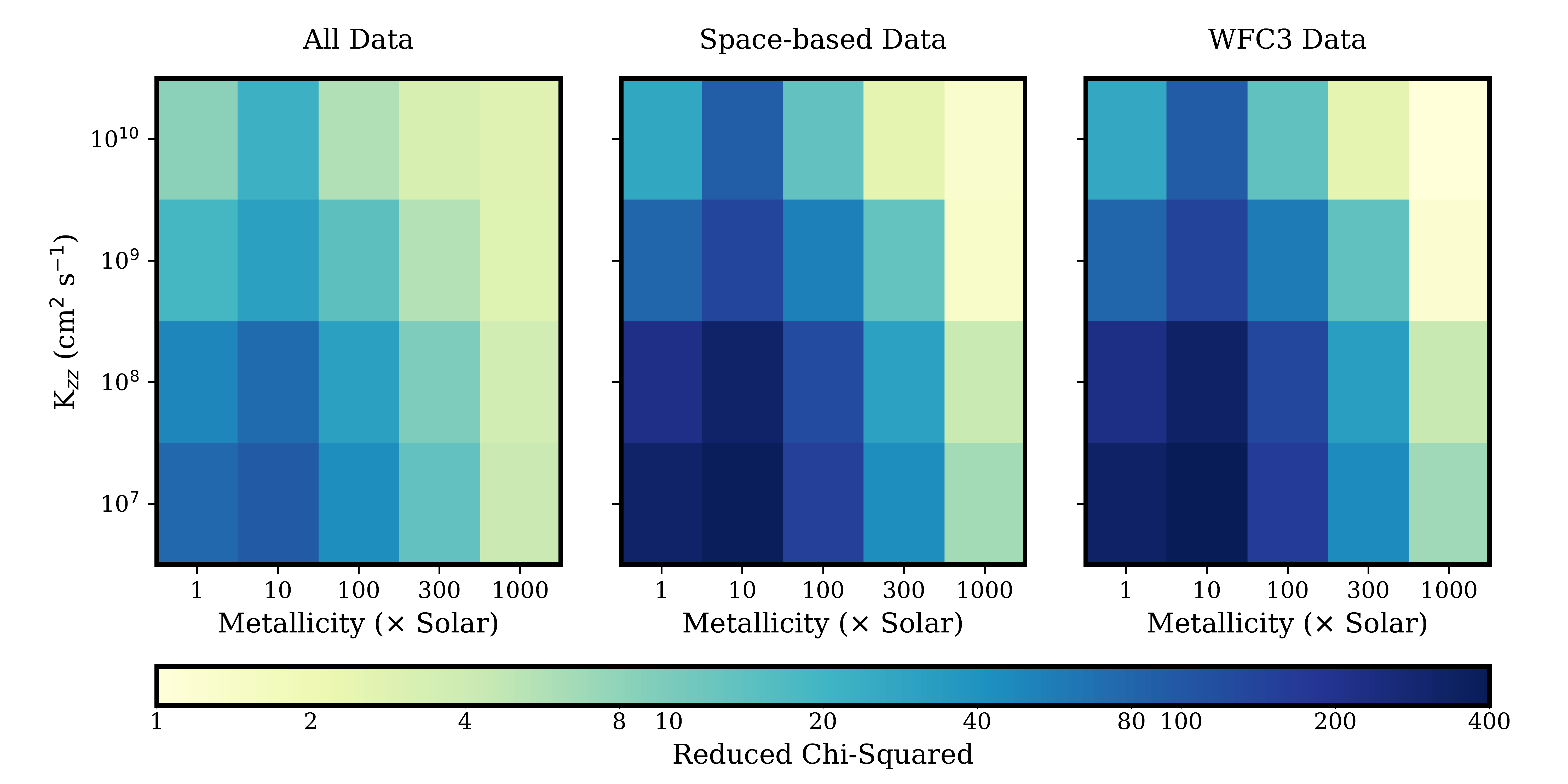}
\caption{The reduced chi squared of fits of our synthetic transmission spectra to all observed GJ 1214 b transmission spectra (left), just space-based observations (middle), and just \citet{kreidberg2014} WFC3 observations (right). }
\label{fig:redchisqr}
\end{figure}

The available data ranges from 0.4 to 4.5 $\mu$m, but the large scatter in the data shortward of 1 $\mu$m, which is significantly greater than the difference between our models, severely impacts our evaluation of the goodness of fit and the validity of our models. This scatter in the data is likely due to uncorrected systematics or stellar effects \citep[e.g.][]{rackham2017}. Therefore, we also consider a fit to observations made only by space telescopes (HST and Spitzer), and a fit to only the observations of \citet{kreidberg2014}, which have the highest precision out of all the data. Figure \ref{fig:compobsspacewfc3} shows these alternative fits, while Figure \ref{fig:redchisqr} and Table \ref{table:redchisqrdtable} show the reduced chi squared of all fits, compared with that of a flat line.

The higher \kzz, higher metallicity simulations are in better agreement with the data, which is expected since a higher \kzz results in more cloud particles being lofted to the high altitudes probed by transmission spectroscopy, while higher metallicities increase cloud mass and decrease the magnitude of spectral features. In particular, the 1000 $\times$ solar metallicity, \kzz= 10$^{10}$ cm$^2$ s$^{-1}$ case reaches reduced chi-squared values ($\sim$ 1) lower than that of the flat line case. The significance of this difference in goodness of fit can be quantified by comparing the Bayesian Information Criterion \citep[BIC;][]{schwarz1978},

\begin{equation}
\label{eq:bic}
{\rm BIC} = n_p \ln{n_d} - 2\ln{\mathcal{L}}
\end{equation}

\noindent where $n_d$ is the number of data points, $n_p$ is the number of free parameters, and $2\ln{\mathcal{L}}$ is the chi-squared. Because we are comparing only 1 instance of our model and not the collective model grid, the base radius is the only free parameter for both the best fit model and the flat line. Since $n_d$ is also the same between these two models, BIC reduces to just the difference in chi-squared. For the WFC3-only data, the difference in chi-squared is only 0.3, which is insignificant \citep{kass1995}. However, when the two Spitzer points are included, the difference in chi-squared increases to 8.4, which is compelling. Close inspection of Figure \ref{fig:compobsspacewfc3} suggests that our best fit model fits the 4.5 Spitzer point better than a flat line. In our model, the 4.5 micron point is slightly elevated above the level of the 3.6 micron Spitzer point and the HST WFC3 points due to CO absorption. 

\begin{deluxetable}{lcccc}
\tablecolumns{5}
\tablecaption{Reduced chi squared of model fits \label{table:redchisqrdtable}}
\tablehead{
\colhead{K$_{zz}$ (cm$^{2}$ s$^{-1}$)} & \colhead{Metallicity ($\times$ solar)} & \colhead{$\chi_{red}^2$ (All)} & \colhead{$\chi_{red}^2$ (Space)} & \colhead{$\chi_{red}^2$ (WFC3)}}
\startdata
\hspace{8mm}$10^7$   &	1       &   76.886  &   317.693 &   336.930 \\
\hspace{8mm}$10^8$   &	1       &	49.233  &   214.530 &   224.825 \\
\hspace{8mm}$10^9$   &	1       &	19.079  &   78.226  &   79.081  \\
\hspace{8mm}$10^{10}$&  1       &	8.406   &   27.108  &   26.047	\\
\hspace{8mm}$10^7$	 &	10      &	95.644  &   373.804 &   399.559	\\
\hspace{8mm}$10^8$	 &	10      &	72.611  &   317.023 &   337.707	\\
\hspace{8mm}$10^9$	 &	10      &	30.718  &   134.464 &   140.274 \\
\hspace{8mm}$10^{10}$&	10      &	21.522  &   90.131  &   92.979  \\
\hspace{8mm}$10^7$	 &	100     &	43.976  &   151.819 &   162.508	\\
\hspace{8mm}$10^8$	 &	100     &	30.559  &   122.314 &   130.559	\\
\hspace{8mm}$10^9$	 &	100     &	14.355  &   54.607  &   57.481	\\
\hspace{8mm}$10^{10}$&	100     &	5.573   &   13.347  &   13.533	\\
\hspace{8mm}$10^7$	 &	300     &	13.224  &   43.324  &   46.359	\\
\hspace{8mm}$10^8$	 &	300     &	9.471   &   30.130  &   32.097	\\
\hspace{8mm}$10^9$	 &	300     &	5.504   &   12.840  &   13.477	\\
\hspace{8mm}$10^{10}$&	300     &	3.154   &   2.480   &   2.479	\\
\hspace{8mm}$10^7$	 &	1000    &	4.154   &   6.373   &   6.755	\\
\hspace{8mm}$10^8$	 &  1000    &	3.587   &   4.193   &   4.351	\\
\hspace{8mm}$10^9$	 &	1000    &	2.838   &   1.383   &   1.191	\\
\hspace{8mm}$10^{10}$&	1000    &	2.778   &   1.267   &   1.010   \\
\hline
\multicolumn{2}{c}{Flat line} & 2.810 & 1.634 & 1.022 \\
\enddata
\end{deluxetable}

\begin{figure}[hbt!]
\centering
\includegraphics[width=1.0 \textwidth]{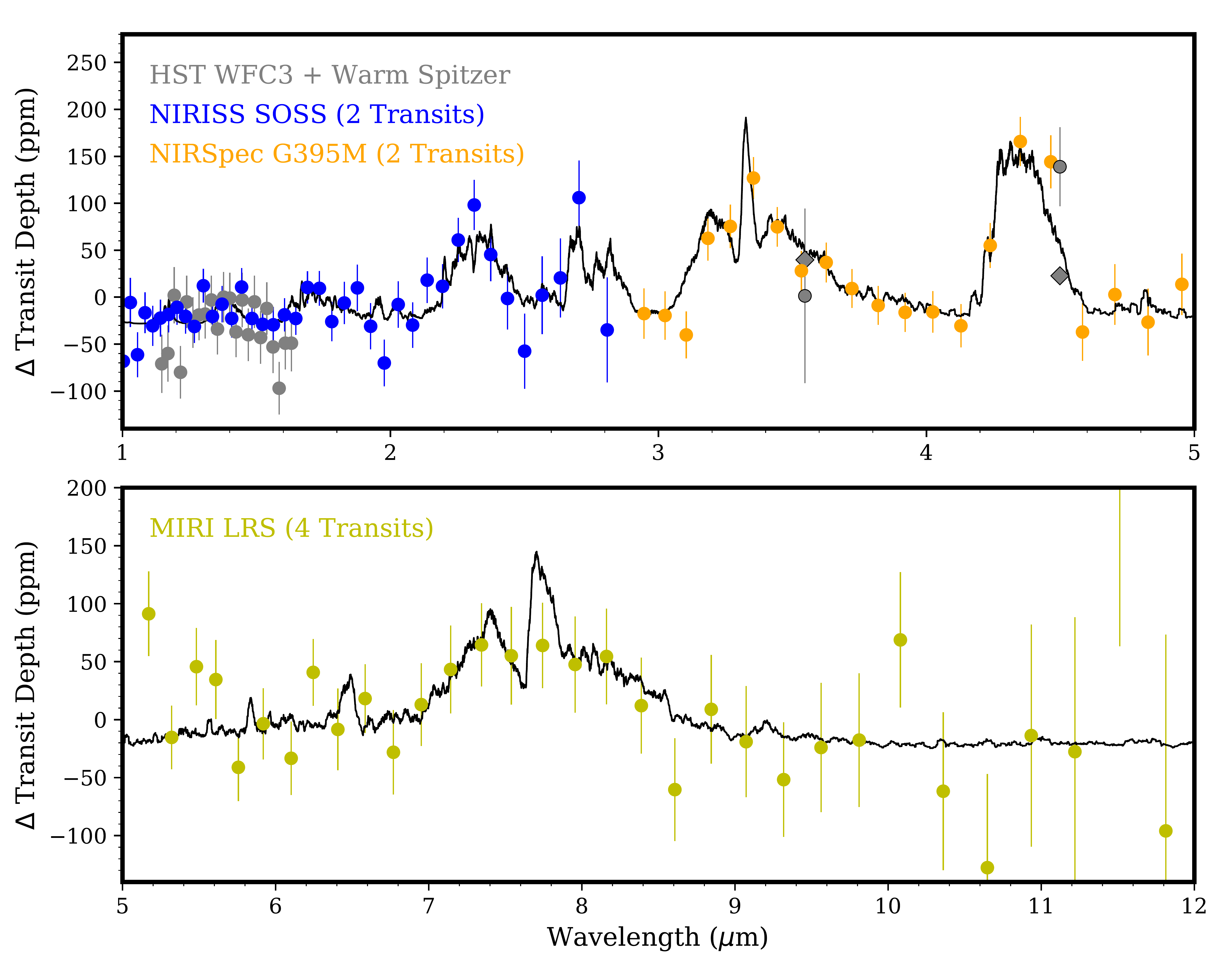}
\caption{Comparison of the 1000 $\times$ solar metallicity, \kzz= 10$^{10}$ cm$^2$ s$^{-1}$ model transmission spectrum (black) with existing HST WFC3 \citep{kreidberg2014} and Spitzer \citep{gillon2014} observations (gray) and synthetic JWST observations (top: NIRISS SOSS in blue, NIRSpec G395M in orange; bottom: MIRI LRS in yellow) generated using Pandexo \citep{batalha2017}. }
\label{fig:jwstpredict}
\end{figure}

Observations with higher spectral resolution covering the 4.5 micron region, such as those possible with the James Webb Space Telescope (JWST), will be necessary to better evaluate the cause of the deviation of the 4.5 micron Spitzer point from the mean. As such, we use our best fit model and the PandExo tool \citet{batalha2017} to predict the observability of the existing molecular features. Figure \ref{fig:jwstpredict} shows two sets of possible observations, each assuming 4 transits. At 1--5 microns, two transits for each of NIRISS SOSS and the NIRSpec G395M grism allows for possible identification of methane at $\sim$2.3 $\mu$m and 3.3 $\mu$m and CO at 4.3 $\mu$m. Beyond 5 microns, 4 transits for MIRI LRS can lead to possible detection of water and methane between 7 and 8.5 microns. These observations can potentially help constrain the atmospheric C/O ratio. Quantifying how well the abundances of these molecules could be constrained by JWST would require conducting a retrieval on the synthetic data, which is beyond the scope of this paper and is left for a future publication.

\subsection{Mixed Clouds}\label{sec:mixedclouds}

As discussed in ${\S}$\ref{sec:micro}, there are significant unknowns regarding how ZnS may form clouds in exoplanet atmospheres. In this work we assume that Zn adsorbs on pure KCl cloud particles, where it reacts with H$_2$S to form ZnS, which then persists on the KCl particle as an enveloping layer surrounding the KCl core. Due to the uncertainties in how Zn may interact with KCl, we do not investigate how ZnS clouds vary with changes in \kzz and metallicity, as we have done for KCl. Instead, we focus on the 1000 $\times$ solar metallicity, \kzz= 10$^{10}$ cm$^2$ s$^{-1}$ case and vary a key variable in the ZnS nucleation rate: the desorption energy E$_{\rm des}$, which we allow to be 0.1 and 1 eV. This exercise allows for first time evaluation of how uncertainties in this microphysical parameter, which can be measured in the laboratory, affects exoplanet cloud distributions.

\begin{figure}[hbt!]
\centering
\includegraphics[width=0.8 \textwidth]{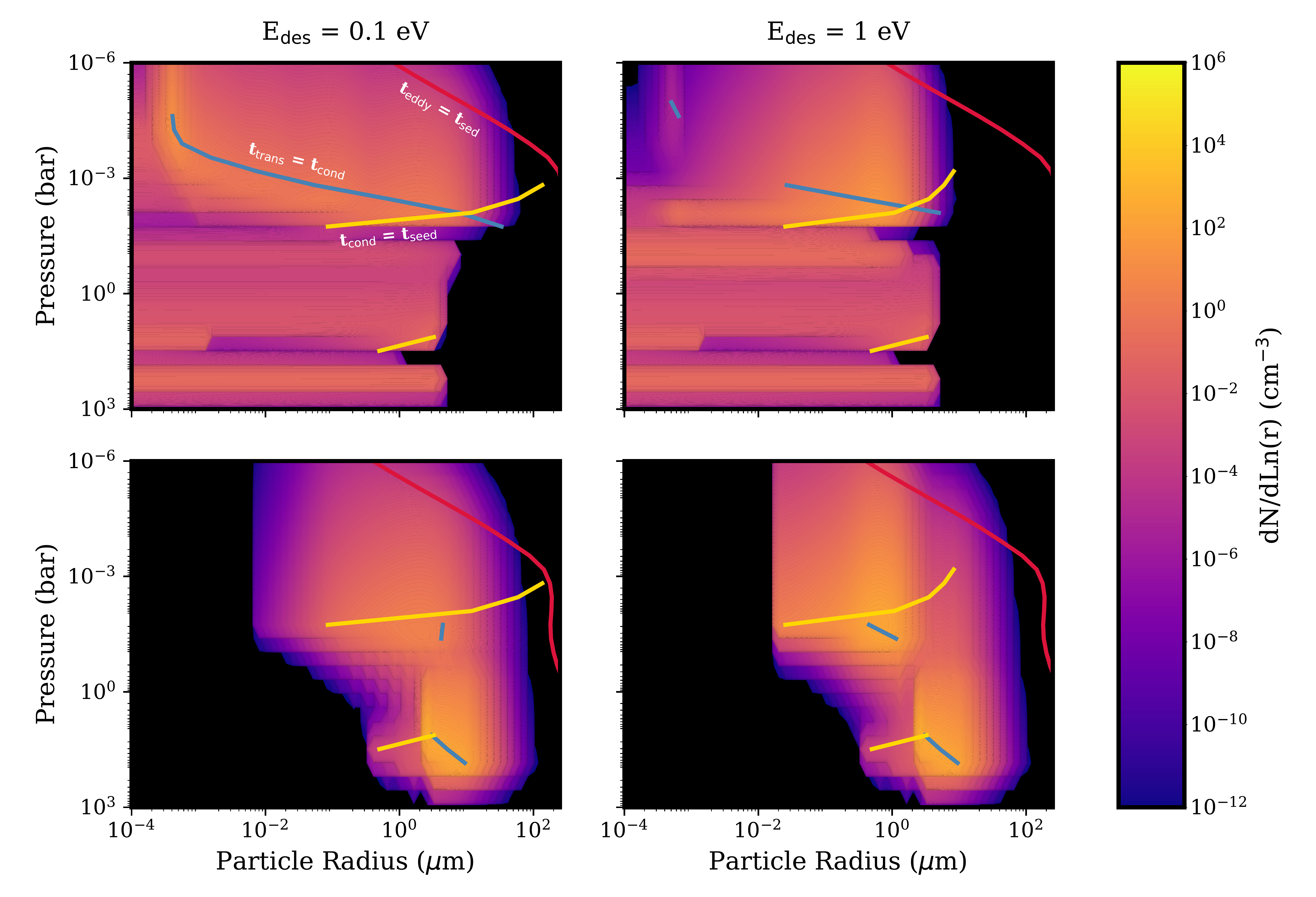}
\caption{Cloud particle number density as a function of particle radius and atmospheric pressure level for pure KCl (top) and mixed (bottom) particles for the E$_{\rm des}$ = 0.1 eV (left) and 1 eV (right) cases. 1000 $\times$ solar metallicity a \kzz value of 10$^{10}$ cm$^{2}$ s$^{-1}$ are assumed. The blue curves show where the timescale of growth by condensation equals the transport timescale; the red curves show where the eddy diffusion timescale equals the sedimentation timescale; and the yellow curves show where the timescale of growth by condensation of KCl particles equals the timescale of loss of KCl particles to ZnS nucleation.}
\label{fig:mixcontour}
\end{figure}

Allowing ZnS to heterogeneously nucleate on KCl results in the formation of a significant mixed cloud distribution and changes to the pure KCl cloud distribution, as shown in Figure \ref{fig:mixcontour}. As with the pure KCl case, the cloud distribution as a function of particle radius and atmospheric pressure level can be understood by appealing to timescales. Here we define a new timescale $t_{seed}$, which is the timescale for the nucleation of ZnS on KCl cloud particles. $t_{seed}$ joins $t_{cond}$ and $t_{trans}$ in defining the KCl and mixed cloud distributions by limiting the growth of KCl past a certain size, as heterogeneous nucleation tends to favor large CNs on account of the Kelvin effect. For example, in the E$_{\rm des}$ = 0.1 eV case, the $t_{cond}$ = $t_{trans}$ curve for pure KCl once again define the peak particle number density, as smaller particles quickly grow to that radius, while larger particles are controlled by transport (eddy diffusion in this case). However, when the $t_{cond}$ = $t_{trans}$ curve meets the $t_{cond}$ = $t_{seed}$ curve, KCl particles can no longer continue growing, as they are more likely to be nucleated upon and form cores in mixed ZnS/KCl particles. Therefore, the junction of the $t_{cond}$ = $t_{trans}$ and $t_{cond}$ = $t_{seed}$ curves define the maximum size of pure KCl particles. At 10 bars the $t_{cond}$ = $t_{seed}$ curve appears again without the $t_{cond}$ = $t_{trans}$ curve, suggesting that all KCl particles there grow quickly until they are enveloped by nucleating ZnS. Evaporation dominates presure levels immediately above and below 10 bars.

Once a mixed cloud particle forms, they follow the timescales associated with mixed particles. Specifically, particles ``emerging'' from the $t_{cond}$ = $t_{seed}$ curve in the upper cloud deck (above 1 bar) exist in an area of phase space where $t_{cond}$ $>$ $t_{trans}$, and thus the mixed cloud is almost entirely controlled by transport, which is dominated by eddy diffusion up to the $t_{eddy}$ = $t_{sed}$ curve. A small section of the $t_{cond}$ = $t_{trans}$ curve exists at a few tens of mbar, but its effects are not significant. In the lower cloud deck, mixed particles from the $t_{cond}$ = $t_{seed}$ curve can grow to the $t_{cond}$ = $t_{trans}$ curve before being transported via eddy diffusion, forming the particle number density peak at a few microns at and below 1 bar.

The E$_{\rm des}$ = 1 eV case is qualitatively similar, though two significant differences exist. First, the pure KCl cloud is controlled entirely by transport for particle radii between 1 and $\sim$20 nm above 1 mbar, i.e. no significant nucleation or growth by condensation occurs here. This may be due to decreased homogeneous nucleation rates of pure KCl arising from lower KCl vapor supply from depth, as KCl cores of mixed particles are more numerous. Second, the mixed particle $t_{cond}$ = $t_{trans}$ curve defines a visible particle number density maximum, which could be due to the increased abundance of mixed particles from the increased heterogeneous nucleation rates.

\begin{figure}[hbt!]
\centering
\includegraphics[width=0.6 \textwidth]{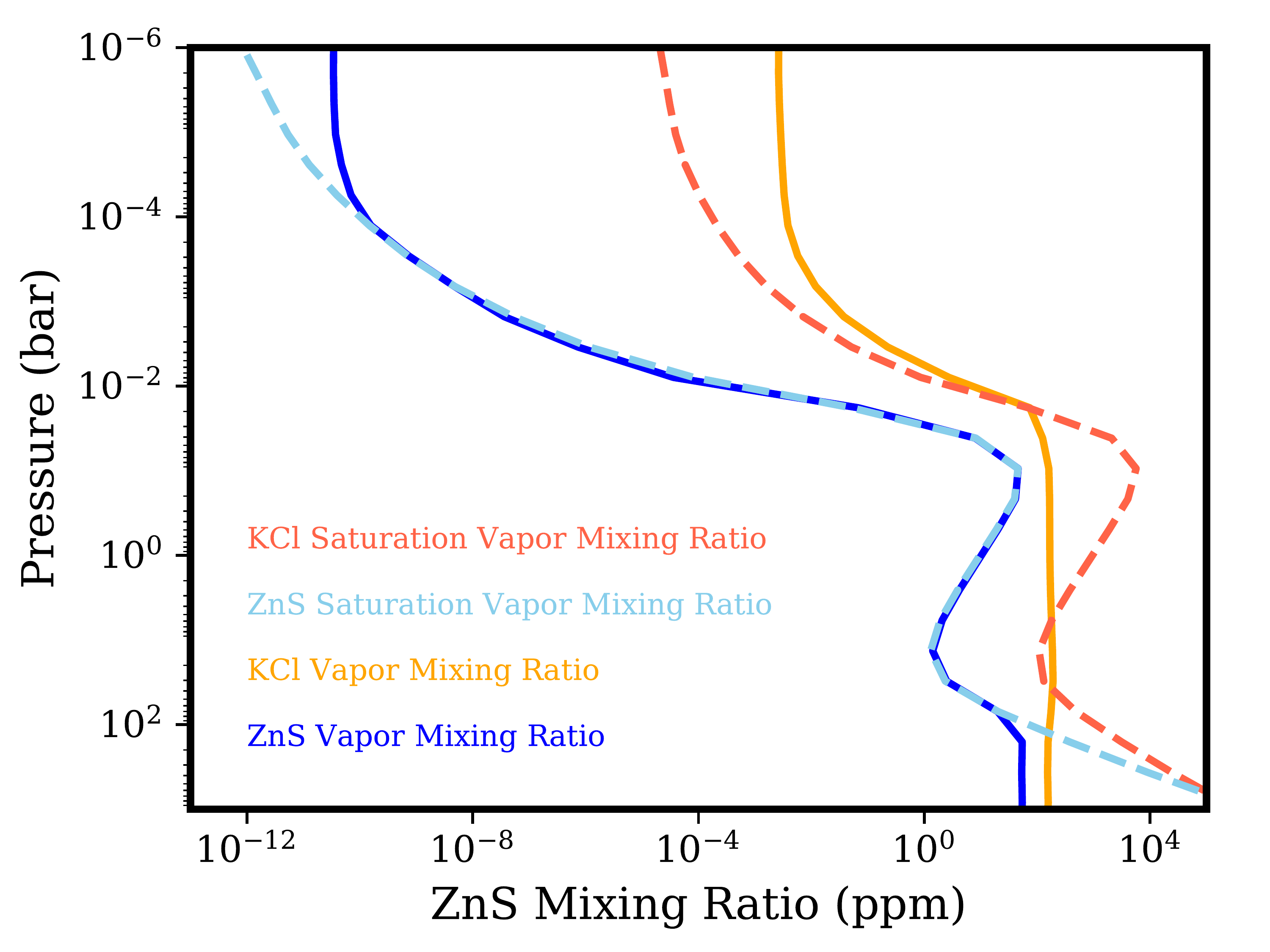}
\caption{ZnS (blue) and KCl (orange) vapor mixing ratio for the 1000 $\times$ solar metallicity, \kzz= 10$^{10}$ cm$^{2}$ s$^{-1}$, and E$_{\rm des}$ = 1 eV case, compared to the ZnS (light blue, dashed) and KCl (red, dashed) saturation vapor mixing ratios.}
\label{fig:mixzns}
\end{figure}

Heterogeneous nucleation depletes Zn vapor to saturation at altitudes above 100 bars, which defines the base of the lower cloud deck (Figure \ref{fig:mixzns}). However, because ZnS's saturation vapor pressure is dependent on metallicity, at high metallicity ZnS is actually less volatile than KCl. Therefore, below 10 mbars, which is the KCl cloud base, the core of the mixed particles should evaporate. How this manifests microphysically is uncertain. If KCl vapor can escape the ZnS shell without destroying the mixed particle, or if the KCl is not bound in a core but rather makes up some other part of a composite particle (i.e. dirty grain), then such particles may survive down to 100 bars, when ZnS thermally decomposes. If the vapor escapes catastrophically then such particles may be shattered at the KCl cloud base.

\begin{figure}[hbt!]
\centering
\includegraphics[width=0.6 \textwidth]{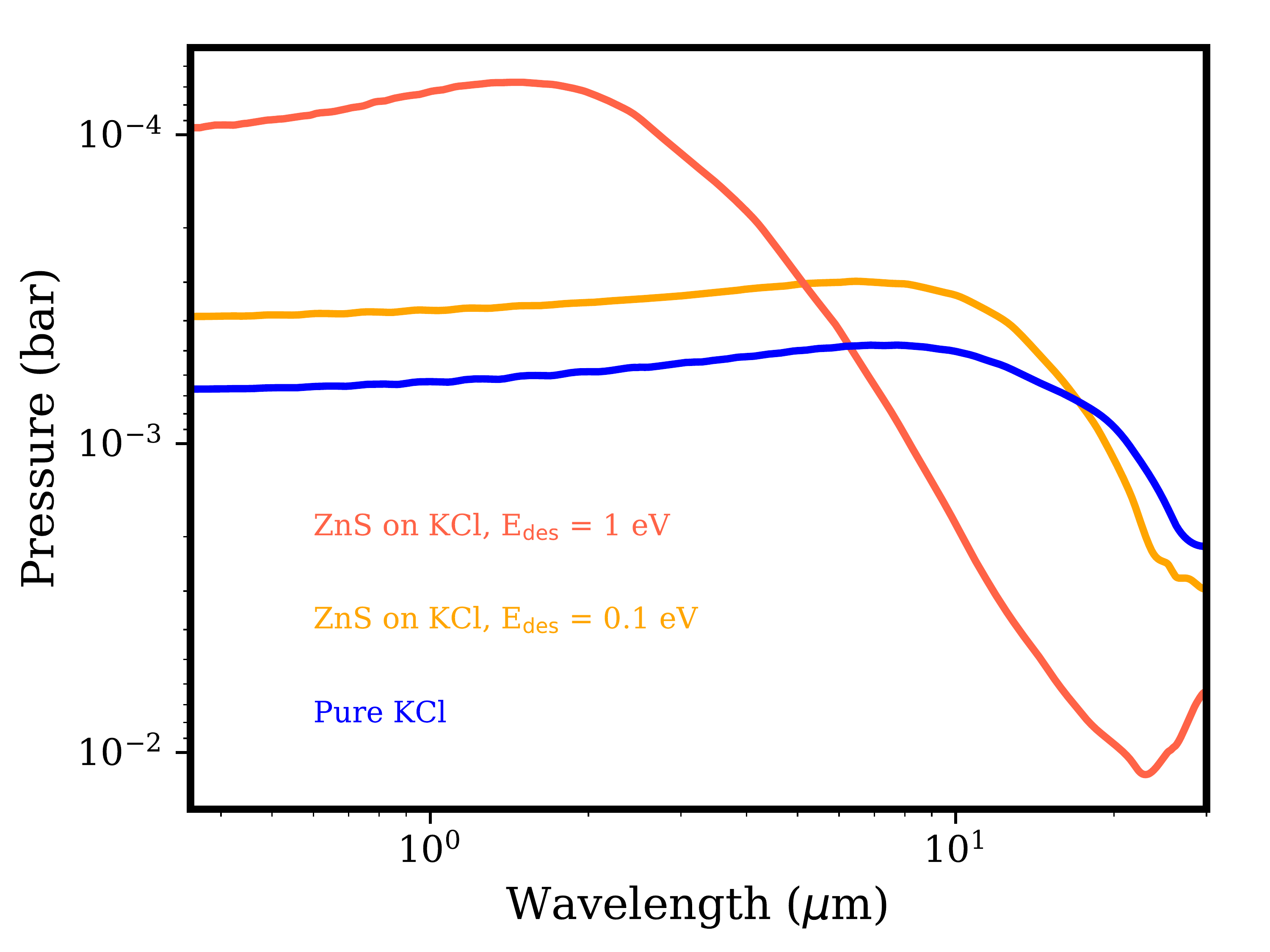}
\caption{Pressure levels where the cloud optical depth equals 1 for the E$_{\rm des}$ = 1 eV (red) and E$_{\rm des}$ = 0.1 eV (yellow) mixed cloud cases, and the pure KCl cloud case (blue), assuming \kzz= 10$^{10}$ cm$^{2}$ s$^{-1}$ and an 1000 $\times$ solar metallicity atmosphere.}
\label{fig:mixtau}
\end{figure}

The addition of mixed clouds alters the total particle distribution considerably. In particular, the effective particle radius is reduced due to inefficient condensational growth, leading to increased cloud opacity at short wavelengths and decreased cloud opacity at longer wavelengths (Figure \ref{fig:mixtau}). However, the particles are still sufficiently large that no ZnS cloud spectral features appear \citep{morley2015,charnay2015b}. In addition, the shape of the $\tau$ = 1 curve for the E$_{\rm des}$ = 1 eV case would appear to not match the observations due to the strong opacity fall off longward of $\sim$3 $\mu$m, though it's possible that mixed clouds simulated using other combinations of \kzz and metallicity could attain similar opacities as that required by the observations while using the same E$_{\rm des}$.

\section{Discussion}\label{sec:discussion5}

\subsection{Comparison with Previous Works}\label{sec:compprev}

\subsubsection{\citet{morley2015}}

\begin{figure}[hbt!]
\centering
\includegraphics[width=0.8 \textwidth]{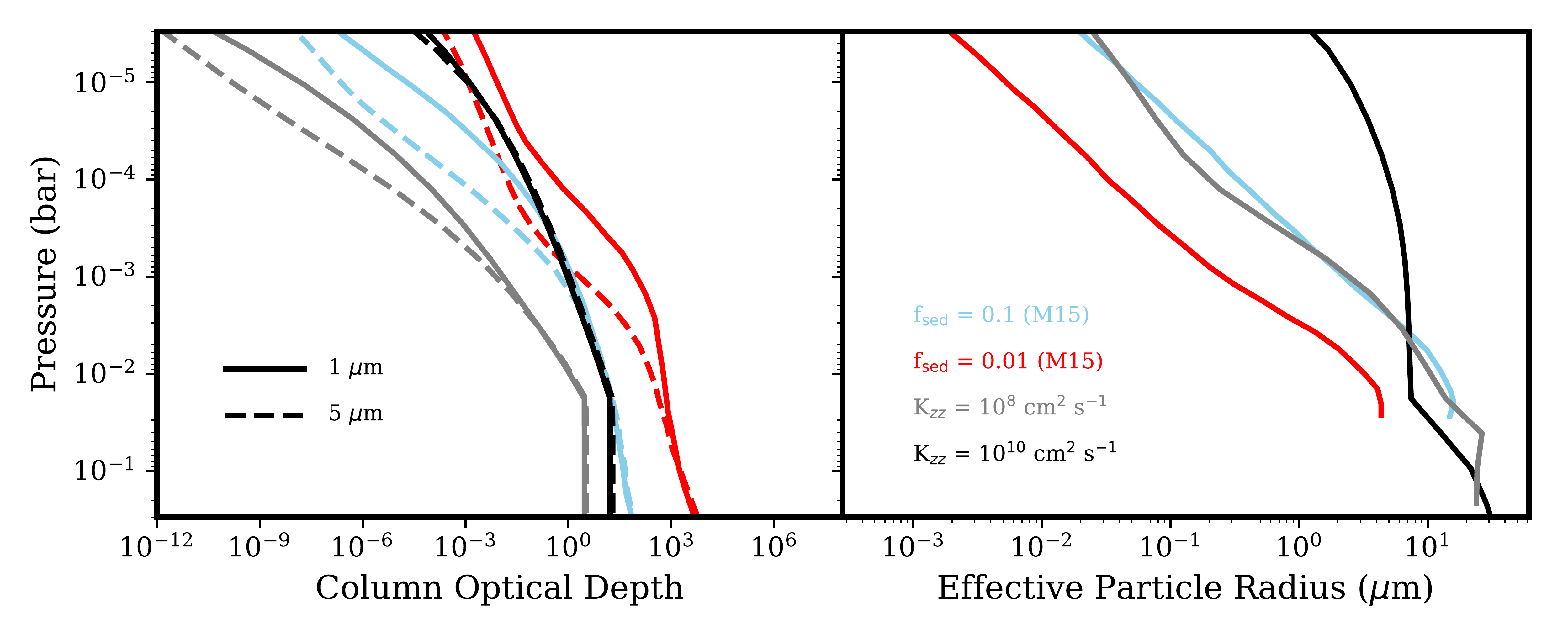}
\caption{Comparison of the computed total cloud column optical depth (left) and effective particle radius (right) profiles between \citet{morley2015} and this work. A 300 $\times$ solar metallicity atmosphere is considered. f$_{\rm sed}$ = 0.1 (blue) and 0.01 (red) cases from \citet{morley2015} are compared to our \kzz= 10$^8$ (gray) and 10$^{10}$ cm$^2$ s$^{-1}$ (black) cases. For the column optical depth profiles, solid lines are the profiles at a wavelength of 1 $\mu$m, while dashed lines are those at 5 $\mu$m.}
\label{fig:morleycomp}
\end{figure}

\citet{morley2015} used a radiative--convective model to generate simulated GJ 1214b atmospheres with 100--1000 $\times$ solar metallicity. The cloud distribution was generated using the \citet{ackerman2001} equilibrium cloud model, which computes cloud mass mixing ratios and particle sizes by balancing particle sedimentation with lofting caused by eddy mixing. Microphysical processes, such as nucleation, condensation, and evaporation are not treated. ZnS and Na$_2$S clouds were considered in addition to KCl clouds. The \citet{ackerman2001} cloud model calculates \kzz using mixing length theory in the convective regions of the atmosphere and assumes a small convective heat flux in the radiative regions such that \kzz is nonzero at all pressure levels. The \kzz values calculated vary with pressure between 3 $\times$ 10$^7$ and 3 $\times$ 10$^9$ cm$^2$ s$^{-1}$ (see their Figure 3), within the range we consider in this work but lower than the \kzz we needed to fit the observations. The cloud particle size distribution was assumed to be lognormal. The vertical extent of the clouds is controlled by the sedimentation efficiency parameter, f$_{\rm sed}$; high f$_{\rm sed}$ clouds are flatter and are composed of large particles, while low f$_{\rm sed}$ clouds are more vertically extended and are composed of small particles. In other words, f$_{\rm sed}$ parameterizes microphysical processes, and how well it does so can be evaluated by comparing the column optical depth and effective particle radius profiles, which are shown in Figure \ref{fig:morleycomp} for 300 $\times$ solar metallicity cases.

The microphysical cloud distribution arising from a \kzz value of 10$^8$ cm$^2$ s$^{-1}$, similar to those used by \citet{morley2015}, is capable of reproducing the effective particle radius profile for their f$_{\rm sed}$ = 0.1 case, but underestimates the column optical depth by $\sim$2 orders of magnitude. This is at least partly caused by the lack of ZnS and Na$_2$S cloud opacity in our model. Meanwhile, the column optical depth profile of our \kzz= 10$^{10}$ cm$^2$ s$^{-1}$ case is similar to that of \citet{morley2015}'s f$_{\rm sed}$ = 0.01 case, but our cloud particles are much larger, which results from significant lofting from eddy diffusion. As such, our cloud opacities are much more constant between wavelengths of 1 and 5 $\mu$m, allowing us to fit both the \citet{kreidberg2014} HST WFC3 data and the \citet{gillon2014} Spitzer points. \citet{morley2015} used an f$_{\rm sed}$ value of 0.01 and a metallicity of 1000 $\times$ solar to fit the observations from \citet{kreidberg2014}, but did not consider the Spitzer points.

\subsubsection{\citet{charnay2015b}}


\citet{charnay2015b} used the 3D Generic LMDZ GCM to simulate GJ 1214b's atmosphere assuming 100 $\times$ solar metallicity. They considered the condensation, evaporation, transport, and sedimentation of pure KCl and ZnS cloud particles, which were distributed lognormally with the peak radius being a free parameter. Condensation and evaporation were treated as instantaneous; any condensate vapor that is supersaturated is assumed to immediately condense into particles of the chosen size, and any particles located in subsaturated regions of the atmosphere immediately evaporate into equal-mass vapor (B. Charnay, personal communication). No \kzz's were used since the cloud particles were able to be transported in three dimensions by the GCM. They found that 0.5 $\mu$m particles for both KCl and ZnS obtained the best fits to the \citet{kreidberg2014} HST WFC3 data. However, because of the relatively small particle size, the resulting simulated transmission spectra show considerable molecular spectral features for wavelengths $>$2 $\mu$m with amplitudes of several hundred ppm. The \citet{gillon2014} Spitzer data points were not considered. By comparison, the 100 $\times$ solar metallicity cases in our study exhibit cloud particle radii of 1--10 $\mu$m at 1 mbar, which are caused by high condensational growth rates at low \kzz and vertical lofting at high \kzz.

\subsubsection{\citet{ohno2018}}

\begin{figure}[hbt!]
\centering
\includegraphics[width=0.8 \textwidth]{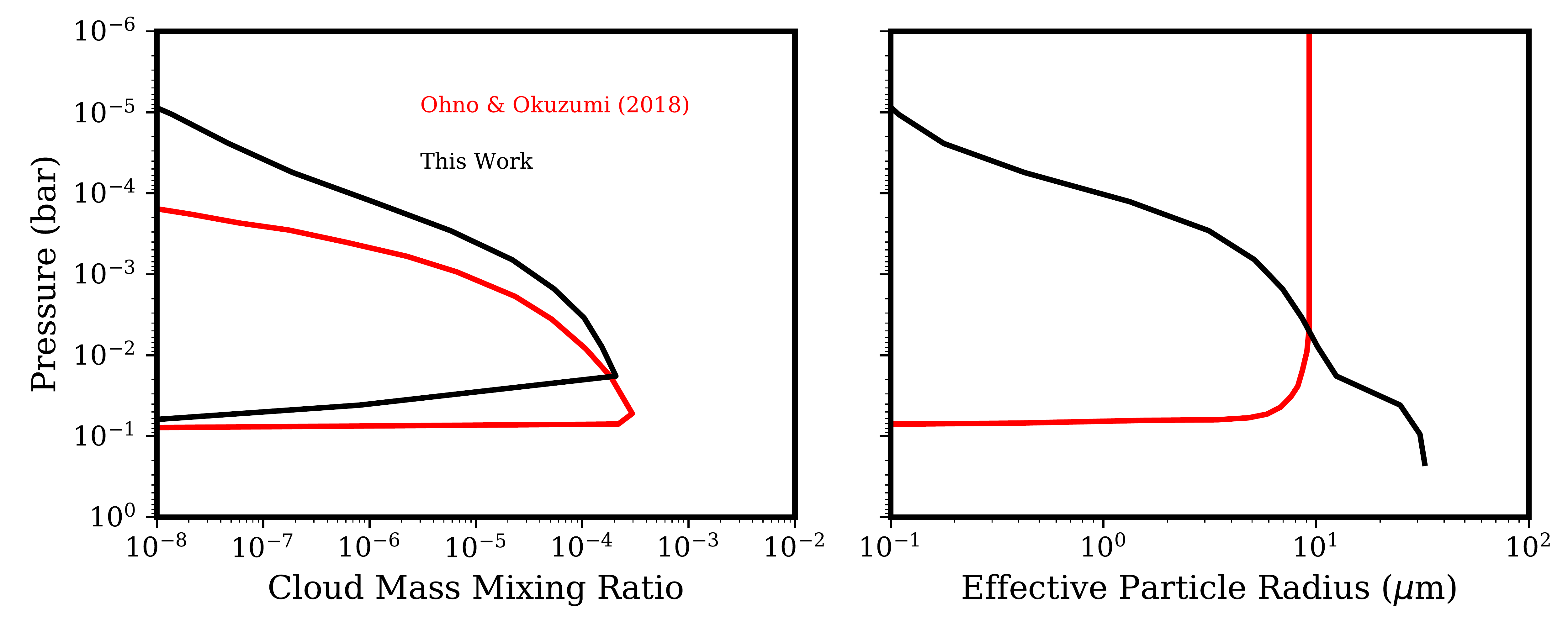}
\caption{Comparison of the computed cloud mass mixing ratio (left) and effective particle radius (right) profiles between \citet{ohno2018} and this work. A 100 $\times$ solar metallicity atmosphere is considered. The n$_{\rm ccn}$ = 1 cm$^{-3}$ (red) case from \citet{ohno2018} is compared to our \kzz= 10$^9$ (black) case.}
\label{fig:ohnocomp}
\end{figure}

The work of \citet{ohno2018} is similar to ours in that they also considered cloud microphysics of KCl clouds in their model and explore a range of metallicities from 1 $\times$ solar to a steam atmosphere for GJ 1214b, though they treat the particle size distribution as narrowly peaked around a computed value rather than as a binned distribution. They use the \kzz profiles from \citet{charnay2015a}, which are at the low end of the range we consider. One key difference between our works is the treatment of nucleation. In our work, nucleation rates are explicitly calculated, while in \citet{ohno2018} CNs composed of KCl with a radius of 1 nm and varying number densities are placed at the cloud base. The radius of 1 nm is consistent with the size of particles produced by nucleation in our model (Figure \ref{fig:microtime}), while the number densities are more difficult to compare since the number density of the ``nucleation mode'' in our size distributions are impacted by the growth rate of these particles to larger sizes. However, for sufficiently high \kzz growth is slower than transport, and so those cases can be compared. For example, number densities $\sim$1--10 cm$^{-3}$ are reached for the nucleation mode in the 100 $\times$ solar metallicity, \kzz= 10$^{9}$ cm$^2$ s$^{-1}$ case, which places this case in the condensation--controlled regime of \citet{ohno2018} (see their Figure 5). The cloud mass mixing ratio profile of our aforementioned case compares favorably with their 100 $\times$ solar metallicity, n$_{\rm ccn}$ = 1 cm$^{-3}$ case (Figure \ref{fig:ohnocomp}, left). The discrepancy in the \kzz used in this comparison stems from the relatively high \kzz required for high rates of homogeneous nucleation due to the necessary supply of vapor from depth. In other words, for the \kzz profiles used in \citet{ohno2018}, the CN number densities that would be generated from homogeneous nucleation is much smaller than those considered in their work. In addition, while \citet{ohno2018} assumed that the CNs only existed at the cloud base, we show that nucleation can occur across multiple scale heights above the cloud base. This explains the difference in the effective particle radius profile between our two works (Figure \ref{fig:ohnocomp}, right): Because their particle formation is restricted to the cloud base and the upper cloud is controlled by transport, their particle sizes are set by conditions at and near the cloud base; in contrast, because nucleation and growth occurs throughout our model atmosphere, the local particle size is controlled more by competing in situ growth and transport rates, such that the effective particle radius increases with increasing pressure in the atmosphere. Despite these differences, however, The cloud top pressures computed by \citet{ohno2018} is qualitatively consistent with our work (Figure \ref{fig:taugrid}), particularly between the lower \kzz and n$_{\rm ccn}$ cases.

\subsection{Beyond Homogeneous Nucleation}\label{sec:beyhom}

Our treatment of cloud microphysics in the atmosphere of GJ 1214b relies on the crucial assumption that KCl clouds form purely through homogeneous nucleation. We did not consider heterogeneous nucleation of KCl clouds because of the uncertainties in the composition, size, and number density of CNs that would require. However, as heterogeneous nucleation is typically preferred in planetary atmospheres in the solar system, its impact on the cloud distribution must be addressed. 

In the event that heterogeneous nucleation supersedes homogeneous nucleation, the cloud distributions that result could be very different from what we have shown here. For example, a large flux of CNs could result in numerous small particles, as the condensate vapor would be distributed among all of them \citep{gao2018,ohno2018}. In other words, the cloud particle number density and size distribution would be dependent on the CNs' number density and size distribution, in addition to their own material properties. 

There are a few possible choices for CNs in exoplanet atmospheres. On Earth, meteoritic dust formed from the condensation of vapor shed by ablating meteorites can potentially act as nucleation sites for high altitude water ice clouds \citep{hunten1980}. If such processes occur in exoplanet atmospheres, then the particle number density of clouds there would be controlled by the flux of meteoritic dust particles, assuming exoplanet cloud materials favor nucleation onto the surfaces of meteoritic dust (i.e. the contact angles between meteoritic dust and cloud condensates are small). This may offer a way to probe the dust environment in exoplanet systems.

Another possible set of nucleation sites are the surfaces of sedimenting photochemical haze particles, which form at high altitudes from photochemistry and polymerization of photolysis products. Such hazes pervade many worlds in the Solar System \citep[e.g.][]{yung1984,zhang2012a,gao2017a}, and similar hazes have been proposed for exoplanets \citep{morley2013,zahnle2016,gao2017b}. As these hazes sediment to deeper atmospheric layers, they can potentially become nucleation sites for supersaturated cloud condensates (again, subject to contact angle considerations). Thus, there may exist a strong link between clouds composed of materials derived from equilibrium chemistry and hazes formed from disequilibrium chemistry.

\subsection{Hazes and High K$_{zz}$'s}\label{sec:hazehighkzz}

Photochemical hazes could also be directly responsible for the flat transmission spectra of GJ 1214 b, as was pointed out by \citet{morley2013,morley2015}. Haze distributions may differ greatly from that of condensation clouds, as they are produced high in the atmosphere rather than being supported by upwelled vapor from below. \citet{kawashima2018} simulated haze formation and evolution on GJ 1214b by accounting for photochemistry, coagulation, diffusion, and sedimentation of haze particles, and found that high rates of photochemical haze formation is necessary to significantly reduce the amplitude of spectral features in the transmission spectra. However, due to the small radius of haze particles, the transmission spectra are heavily sloped for all but the case with the highest considered photochemical production rate, and no comparison to observations were undertaken. 

The observational constraints from the \citet{kreidberg2014} and \citet{gillon2014} data points require relatively large particles (r $>$ 1 $\mu$m) to be situated high in the atmosphere. This poses a conundrum: In order to have large condensational cloud particles at low pressures, \kzz must be large, as was shown in this work. However, such large \kzz values are inconsistent with those computed from GCMs \citep{charnay2015a} by several orders of magnitude, though there exists significant uncertainty in \kzz parameterizations for exoplanet atmospheres. By comparison, haze opacity tends to increase at low \kzz values due to inefficient mixing of haze particles and haze precursors into the deep atmosphere \citep{lavvas2017}. However, haze particles are small and tend to produce scattering slopes in transmission rather than flat ``cloud decks''. Heterogeneous nucleation of condensate vapor onto photochemical haze particles may not solve this problem if the vapor is sufficiently depleted at low pressures due to cloud formation near the cloud base. Sufficiently high haze production rates can produce large particles at low pressures due to higher coagulation rates, but haze production rates extrapolated from the production rates and mixing ratios of simple hydrocarbons and nitriles are highly uncertain \citep{horst2018}. The production rates needed are also several orders of magnitude greater than those expected from extrapolations of Titan haze modeling \citep[e.g.][]{kawashima2018}. One possible solution is the formation of aggregate haze particles from coagulation instead of spherical particles. Such particles are larger in physical dimension, but porous enough that they stay aloft for longer periods of time compared to spheres of the same size. However, aggregate particles has yet to be investigated for exoplanet atmospheres.

\subsection{Coagulation Considerations}\label{sec:coalcon}

\begin{figure}[hbt!]
\centering
\includegraphics[width=0.6 \textwidth]{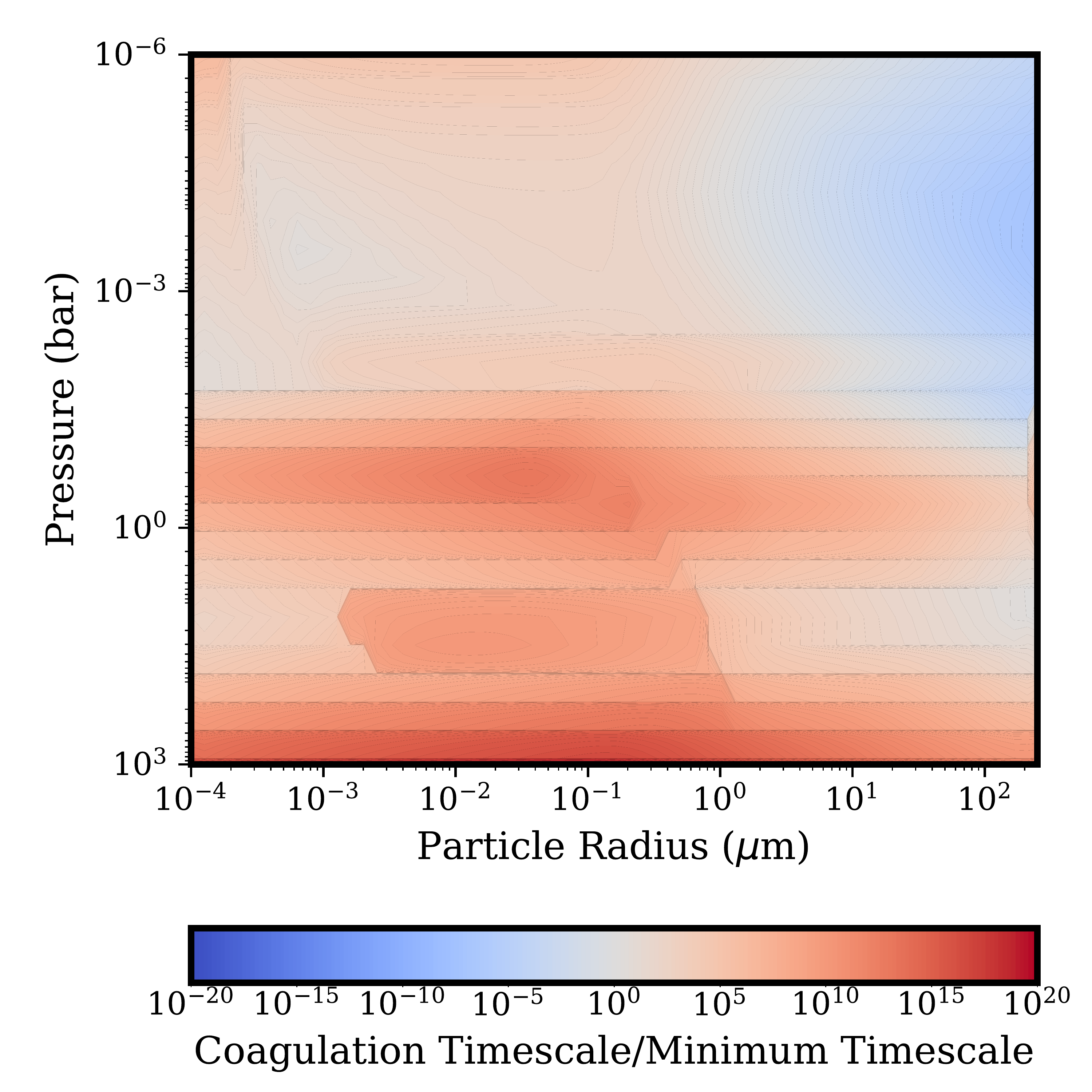} 
\caption{Ratio of coagulation timescale to the minimum of the condensation, nucleation, and transport timescales for the 1000 $\times$ solar metallicity, \kzz= 10$^{10}$ cm$^{2}$ s$^{-1}$ case.}
\label{fig:coagtime}
\end{figure}

\begin{figure}[hbt!]
\centering
\includegraphics[width=0.8 \textwidth]{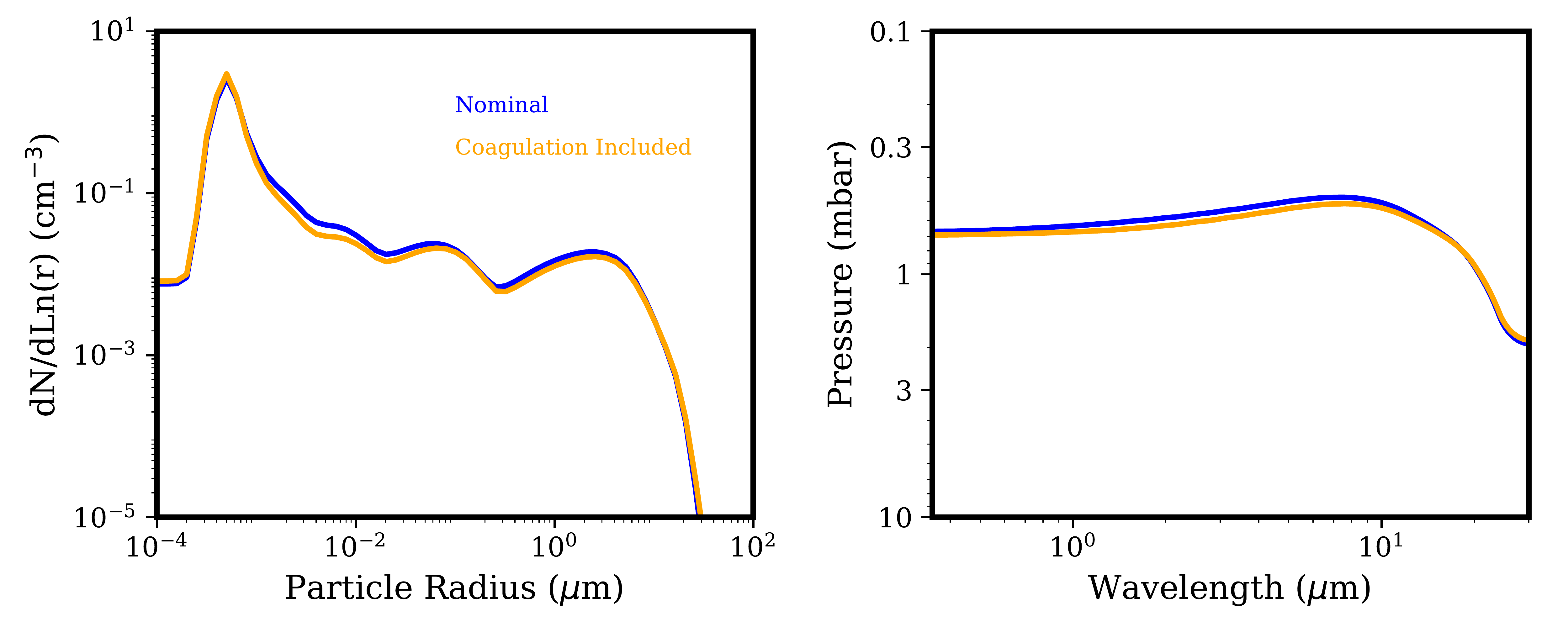}
\caption{Particle size distribution (left) and pressure levels where the cloud optical depth equals 1 (right) for the 1000 $\times$ solar metallicity, \kzz= 10$^{10}$ cm$^{2}$ s$^{-1}$ case with (orange) and without (blue) considering coagulation.}
\label{fig:coagnumtau}
\end{figure}

We have not taken into account coagulation in our work for several reasons. First, it complicates the evaluation of how cloud distributions vary with \kzz and metallicity. Second, the outcome of coagulation in exoplanet atmosphere is not certain: the resulting particles can either be spherical if they are liquid or amorphous, highly porous aggregates if they are solid, or somewhere in between. The particle shape strongly impacts transport rates and opacity, and therefore its uncertainty translates to uncertainty in cloud distributions and atmospheric transmission. However, our results would not be robust if coagulation were an important process. The importance of coagulation can be estimated from timescale analysis. We define a coagulation timescale t$_{coag}$ that is equal to the number density of particles divided by the net rate of change of particle number densities due to coagulation. We compare t$_{coag}$, as defined assuming spherical particles, for the 1000 $\times$ solar metallicity, \kzz= 10$^{10}$ cm$^2$ s$^{-1}$ case to the minimum of all other timescales for the same case in Figure \ref{fig:coagtime}, which shows that coagulation is only important for the largest particles high up in the atmosphere. Here, most of the coagulation occurs via gravitational collection, which is the scavenging of small particles by large particles that fall through the atmosphere faster. We also test the sensitivity of our results to coagulation by turning on the coagulation computations within CARMA and rerunning the same case. The results shown in Figure \ref{fig:coagnumtau} reveal that, even though coagulation is important for part of the phase space, the impact of coagulation is minimal. This results from the negative feedback inherent in coagulation, where high coagulation rates quickly reduce the particle number density enough that coagulation becomes inefficient. Our results are consistent with those of \citet{ohno2018}, who showed that control of particle size by coagulation only occurs for CN number densities far larger than what results from our simulations.

\subsection{Experimental Investigations}\label{sec:labwork}

There remain significant uncertainties in the actual processes at work that produce exoplanet aerosols, some of which could be illuminated by experimental work. In this study, we have focused on homogeneous nucleation of pure KCl and heterogeneous nucleation of ZnS on pure KCl particles. Biomass combustion experiments have shown that homogeneous nucleation of $\sim$10 nm radius KCl particles from KCl vapor readily occurs \citep[e.g.][]{strand2014}, and at least one such experiment showed that heterogeneous nucleation of KCl on SiO$_2$ particles was inefficient \citep{sippula2012}. However, the experimental conditions did not reflect that of exoplanet atmospheres. \citet{osada2010} examined $\sim$1 $\mu$m radius ash particles condensed on the inner wall of a melting furnace and found traces of KCl, NaCl, and ZnS, among others. Na$_2$S, another theorized exoplanet cloud species \citep{morley2012}, was not detected despite being predicted by thermochemistry, suggesting that NaCl formed in its place. ZnS nucleation in solution is common for semiconductor studies \citep[e.g.][]{tiemann2005,senthilkumaar2009,cetinkaya2011}, but we did not find any work investigating ZnS formation in the gas phase. In all, while there are no shortage of experiments focusing on some aspect of KCl and ZnS formation, there are none that can be directly applied to exoplanets.

Experiments that take place at the temperatures and compositions of warm, high metallicity exoplanet atmospheres have recently been conducted \citet{horst2018,he2018}, though with the goal of producing photochemical hazes rather than condensation clouds. This proof of concept opens up the exciting possibility of testing the realities of exoplanet clouds. For example, such experiments could measure the homogeneous nucleation and condensation rates and size distributions of KCl cloud particles by introducing KCl vapor into a background gas analogous to that of a warm exoplanet atmosphere. The efficiency of heterogeneously nucleating KCl on meteoric dust or photochemical hazes can be evaluated by allowing KCl vapor to mix with small amounts of silicate dust or organic soot. The formation of condensed ZnS can be investigated by mixing Zn vapor with H$_2$S with and without foreign surfaces of various compositions. Given that H$_2$S may be lost to photolysis in exoplanet atmospheres \citep{zahnle2009,zahnle2016}, the fate of sulfide clouds like ZnS and Na$_2$S could potentially be assessed by varying H$_2$S abundances and introducing trace amounts of chlorine to quantify whether Na condenses as Na$_2$S or NaCl.

\section{Conclusions}\label{sec:conclusions5}

We have investigated the formation and evolution of KCl and ZnS clouds in the atmosphere of GJ 1214 b by taking into account microphysical processes, including homogeneous/heterogeneous nucleation, growth by condensation, loss by evaporation, and transport by sedimentation and eddy diffusion. We varied the eddy diffusivity and atmospheric metallicity to investigate how KCl cloud distributions arising from homogeneous nucleation are affected by these parameters, and assessed which combination of parameters best fit the available data. We also considered the more complicated scenario of ZnS heterogeneously nucleating on KCl to evaluate the impact of mixing cloud species on cloud distributions and observations. We can make the following conclusions from our results:

\begin{itemize}
  \item The high surface energy of ZnS forbids formation of pure ZnS clouds via homogeneous nucleation. ZnS can only condense if a favorable nucleation surface exists, such as cloud particles of other composition, meteoric dust, or photochemical haze particles, in which case the condensed ZnS abundance is beholden to the abundance of the CNs. The contact angle between ZnS and any of these surfaces is not known but is crucial for calculating the rates of nucleation, necessitating laboratory investigations of these materials. 
  \item KCl cloud distributions arising from homogeneous nucleation can be defined in the pressure--particle radius space by timescales of nucleation, condensation, evaporation, eddy diffusion, and sedimentation. Specifically, the peak in the particle size distribution is defined by equating the timescale of nucleation and condensation with that of transport. If transport is dominated by sedimentation, then the aforementioned peak also defines the maximum particle size per pressure level. If transport is dominated by eddy diffusion, then the maximum particle size per pressure level is defined by the maximum particle size at the cloud base, since those particles can be lofted up to the pressure level where the sedimentation time scale equals the eddy diffusion timescale. The largest particle size at the cloud base is defined by equating the condensation and evaporation timescales. The cloud mass increases with increasing \kzz and metallicity, but at sufficiently high \kzz the increase slows down due to downward mixing of cloud particles.
  \item The pressure level at which the cloud optical depth equals one decreases with increasing \kzz and metallicity, and is mostly constant with wavelength shortward of 10 $\mu$m. Cloud optical depth drops at longer wavelengths.
  \item Data constrain the atmospheric metallicity to at least 1000 $\times$ solar, and the \kzz to be at least 10$^{10}$ cm$^{2}$ s$^{-1}$, under the assumption that the observed flatness of the transmission spectrum is caused by homogeneously nucleated KCl cloud particles. However, other cloud formation mechanisms and aerosol compositions are possible. 
  \item The inclusion of heterogeneous nucleation of ZnS on KCl reduces particle sizes, leading to increased cloud opacity at short wavelengths and decreased cloud opacity at long wavelengths. This depends strongly on how ZnS interacts with KCl, if at all, and requires experimental studies.
  \item Given our best fit model spectrum, we predict that JWST observations of GJ 1214b may reveal the presence of methane, CO, and water, allowing for constraints to be placed on atmospheric metallicity and C/O ratio.
\end{itemize}

Our results reveal the complexities involved in determining the distribution of exoplanet clouds. Their dependence on aspects of the background atmosphere and their own material properties not only increases the difficulty in generalizing their behaviors across planetary parameter space, but also strongly ties them to the rest of the atmosphere. Therefore, models that capitalize on those ties, such as ours, will be indispensable in understanding these exotic atmospheres as we face the onslaught of new exoplanet observations from JWST, WFIRST, and ground--based observatories. In addition, the importance of laboratory studies into the properties of proposed exoplanet cloud species cannot be overstated, as they are instrumental in informing models like ours by providing necessary material properties, as well as observations by providing the optical properties of these materials and their mixtures. As investigations of exoplanets focus more on smaller, cooler worlds, insight into the clouds and hazes that are sure to pervade their atmospheres will become increasingly important in understanding them as a whole. 

\acknowledgments

We thank H. Zhang for her incredible and continuing support throughout the duration of this project. We thank E. Barth, V. Hartwick, C. E. Bardeen, X. Zhang, and D. Powell for helpful discussions regarding the model mechanics. We thank Y. L. Yung, H. A. Knutson, M. S. Marley, M. R. Line, C. V. Morley, and V. Parmentier for enlightening discussions and advice. We thank N. Batalha for much needed guidance on working with PandExo. We thank the anonymous reviewer for motivating the extensive rewrites that have vastly improved this work. We thank H. Ngo for helpful discussions on goodness of fit. P. G. acknowledges support from the NASA Postdoctoral Program, the Heising-Simons Foundation, and an NAI Virtual Planetary Laboratory grant from the University of Washington to the Jet Propulsion Laboratory and California Institute of Technology under solicitation NNH12ZDA002C and Cooperative Agreement Number NNA13AA93A. This work was also supported by the Space Telescope Science Institute (STScI) under the Hubble Space Telescope GO-13665 program (PI Benneke).



\vspace{5mm}


\bibliography{references}
\bibliographystyle{aasjournal}

\end{document}